\documentclass[final]{aipproc}
\usepackage{eucal}

\usepackage{graphicx}
\usepackage{bm}
\usepackage{amsmath}
\usepackage{amsfonts}
\usepackage{amssymb}
\usepackage{psfig}

\newcommand{\gaprox}{$ {\raisebox{-.6ex}{{$\stackrel{\textstyle >}{\sim}$}}} $}
\newcommand{\saprox}{$ {\raisebox{-.6ex}{{$\stackrel{\textstyle <}{\sim}$}}} $}
\newcommand{\boldpi}{\mbox{${\bm \pi}$}}
\newcommand{\boldT}{\mbox{${\bm T}$}}
\newcommand{\boldD}{\mbox{${\bm D}$}}

\newcommand{\boldt}{\mbox{${\bm t}$}}
 
\newcommand{\boldepsilon}{\mbox{${\bm \varepsilon}$}}

\newcommand{\boldtau}{\mbox{${\bm \tau}$}}

\def\L{\Lambda}
\def\R{\right} \def\L{\left} \def\Sp{\quad} \def\Sp2{\qquad}

\layoutstyle{6x9}

\begin{document}
\title{Introduction to Effective Field Theories in QCD
\footnote{Notes taken by LJA and DMC on lectures delivered by UvK
at the Pan-American Advanced Studies Institute on 
New States of Matter in Hadronic Interactions, 
January 7-18, 2002, Campos do Jord\~ao, Brazil.
}}

\author{U. van Kolck${}^{\ast}$}
{address={Department of Physics,
University of Arizona, Tucson, AZ 85721},
address={RIKEN-BNL Research
Center, Brookhaven National Laboratory, Upton, NY 11973}}
\author{L.J. Abu-Raddad}{address={Research
Center for Nuclear Physics, Osaka University, 10-1 Mihogaoka, Ibaraki, 
Osaka 567-0047, Japan}}
\author{D.M. Cardamone}{address={Department of Physics,
University of Arizona, Tucson, AZ 85721}}

\begin{abstract}
We present a simple introduction to the techniques of effective field theory
(EFT) 
and their application to QCD. For problems with more than one energy scale, 
the EFT approach is a useful alternative to more traditional 
model-building strategies. 
The most relevant such problem for this discussion is that of 
making contact between QCD and the hadronic phase of matter. 
As a simple example, an EFT calculation of the bound states of 
hydrogen within QED is sketched. 
A more significant demonstration of the power of EFTs, 
the construction of the chiral Lagrangian and chiral perturbation theory, 
is also included. 
The results provide us with the road map to a complete QCD-based theory 
of nuclear matter at nonzero temperatures and densities, 
a vital component to a quantitative understanding of the phase transition
from hadron gas to quark-gluon plasma.
\end{abstract}

\maketitle

\section{Introduction}
Effective field theory (EFT) is a theoretical prescription for constructing 
theories spanning multiple energy scales. 
The physics of a system may appear radically different at different 
energy scales, due to low-energy restrictions on available degrees of 
freedom and symmetries. 
When trying to construct a theory which spans energy scales, traditional 
methods of physics can therefore be difficult to apply. 
Rather than stumbling on this obstacle, however, EFT provides a method to 
use the physical difference between energy regimes to advantage. 
We present an introduction to the basic ideas of this useful technique,
from the perspective of its application to nuclear physics.
The emphasis is pedagogical.
Comprehensive and up-to-date reviews can be found in Ref. \cite{vankolck99}.
For more extensive lectures with a similar perspective, 
the reader should refer to Ref. \cite{phillips02}.

In a typical experiment designed to produce a quark-gluon plasma (QGP), 
two heavy nuclei collide at relativistic speeds. 
The experimental signal that results is necessarily dependent not just 
on the nature of the plasma, but on the nature of the hadronic phase 
as well as the physics of the phase transformation. 
Trying to understand such an experiment with only an understanding 
of the plasma phase, therefore, is as doomed to failure as attempting 
to understand the latent heat of a water-to-steam transformation using 
only the kinetic theory of gases.

To remedy this problem, we must seek a true quantitative understanding 
of the physics of the hadronic phase. The theory we seek should furthermore 
be based on the microscopic theory of QCD. 
The development of theoretically-sound hadron interactions has been 
the major problem of nuclear physics since its inception. 
This theory must be valid over the significant 
temperatures and densities of the phase transformations typical 
in experiment. 
However, even a more limited approach that is restricted
to the hadronic phase can be useful.
One would like, for example, to estimate the position
of the QGP transition line. 
Moreover, it is believed that
the phase diagram of QCD presents other interesting
features such a liquid-gas transition at lower temperatures
and at densities close to that of equilibrium for cold nuclear matter.

At zero baryon density,
an expansion in temperature $T$ of the free energy density of a
pion gas can be derived in EFT and 
reveals the presence of a transition temperature $T_c$
at the point where the expansion diverges \cite{gerber89}.
A more precise estimate
can then be obtained by considering the effects of 
higher-mass mesons \cite{rafelski},
in good agreement with lattice results \cite{lattice}.

At non-zero baryon density $\rho$, on the other hand,
the situation is much more complicated.
Lattice QCD simulations are hampered by the infamous sign problem.
EFT might therefore be the only way to study the problem,
but it requires a conceptual leap from the work with mesons alone,
because nucleons bind.
Perturbative expansions for the energy
density of dilute boson and fermion gases 
are well known: see, e.g., Ref. \cite{many} for $T=0$.
These are expansions in $\rho a^3$,
where $a$ is the scattering length, which
is essentially the scattering amplitude at zero energy.
The problem is that the nucleon-nucleon ($NN$) scattering
length is large, so that  
the corresponding expansion fails 
way before the QGP phase transition.
Indeed, the interesting
physics of bound states such as nuclei is associated with
$\rho^{1/3} \gaprox 1/a$.
The failure of a perturbative expansion in density implies that 
we must effectively resum such an expansion. 

The reader may be somewhat surprised by the apparent failure in this regime 
of the perturbative approach, upon which so much of particle physics is based. 
This circumstance may be attributed to the complexities of the nonzero 
temperature and density requirements, as well as the striking disparities 
of the two energy scales we would like to connect. 
On the one hand, the consensus of the majority of the nuclear-physics 
community holds that in nuclei
\begin{itemize}
\item nucleons are non-relativistic;
\item they interact via essentially two-body forces, with smaller 
contributions from many-body forces;
\item the two-nucleon interaction generally possesses a high degree 
of isospin symmetry;
\item external probes usually interact with mainly one nucleon 
at a time.
\end{itemize}
By contrast, in QCD
\begin{itemize}
\item the $u$ and $d$ quarks are relativistic;
\item the interaction is manifestly multi-body, involving exchange of multiple 
gluons;
\item there is no obvious isospin symmetry;
\item external probes can, and often do, interact with many quarks at once.
\end{itemize}

It should not be surprising, then, that some new ideas are required to merge 
these two extraordinarily different bodies of theory. 
Of course, we expect that QCD encompasses the physics of hadronic interactions.
The root of the problem must therefore lie in the difference of energy scales.

In fact, constructing a QCD-based theory of the hadronic phase is a problem 
which involves three separate energy scales spanning three orders of magnitude.
The first and most obvious to a reader well versed in high-energy physics 
is the typical energy scale of QCD, 
\begin{equation}
M_{QCD} \sim 1 ~\textrm{GeV}. 
\end{equation}
The masses of all hadrons except the pion fall within this scale
\footnote{Here and throughout the paper we use units where $\hbar=c=1$.},
and the scale of chiral symmetry breaking
is thought to be $M_\chi = 4\pi f_\pi$,
where  $f_\pi \simeq 93$ MeV is the pion decay constant.
The second scale, 
\begin{equation}
M_{nuc} \sim 100 ~\textrm{MeV}, 
\end{equation}
represents the typical momentum of nucleons in a nucleus:
the inverse root-mean-square charge radius of
light nuclei, or the Fermi momentum of
equilibrium nuclear matter.
It contains also the 
pion decay constant itself, the mass difference between the 
delta isobar and the nucleon,
and the mass of the pion. 
The final energy scale is the typical energy scale of a nucleus,
\begin{equation}
\frac{M_{nuc}^2}{M_{QCD}} \sim 10 ~\textrm{MeV}. 
\end{equation}
The binding energy per nucleon of a nucleus is typically a few MeV. 
For example, the binding energy of $^4$He is 28.296~MeV, and 
the binding energy per nucleon of infinite nuclear matter
is 16~MeV.

Since the goal is to construct a theory valid over an energy range of 
three orders of magnitude, and spanning regimes in which the physics 
is quite disparate, the problem is somewhat daunting. 
We might hope to seek inspiration from an analogous problem in atomic physics, 
the simple case of a hydrogen atom in the ground state. 
A qualitative argument suffices for this discussion.
We know from experiment that the mass of an electron is
\begin{equation}
m_e \simeq 0.511 ~\textrm{MeV}.
\end{equation}
The Hamiltonian is, to a good approximation, 
$H=p^2/2m_e - \alpha/r$, where 
$\alpha=e^2/4\pi \simeq1/137$ is the fine-structure constant.
{}From the uncertainty principle we know that
the electron momentum is $p \sim 1/R$, if the size of the atom
is $R$. Minimizing $E(R) = 1/2m_e R^2 -\alpha/R$ yields
$R= 1/\alpha m_e$, or
\begin{equation} \label{hbound}
p\sim \alpha m_e =3.6 ~\textrm{keV}.
\end{equation}
Finally, the binding energy $B=-E(1/\alpha m_e)$ is
\begin{equation}\label{hboundbound}
B\sim \frac{1}{2}\alpha^2 m_e=13.6 ~\textrm{eV}.
\end{equation}
This problem, like ours, possesses three distinct energy scales, on
the order of $m_e$, $p$, and $B$. It is easy to see from the above,
however, that in the atomic problem all three scales are coupled by
the small fine-structure constant $\alpha$. It is not so clear what the
analogous coupling constant should be in the nuclear problem. In fact, the lack
of a clear small coupling constant is one of the main difficulties of
nuclear physics. Even though our problem involves QCD, the strong
coupling strength $\alpha_s$ is not useful to us, 
for at the highest energy scale
in the problem $\alpha_s(1\textrm{GeV})\approx 1$
and perturbative QCD, valid at larger energies, breaks down. 

If all obvious coupling parameters are of order one in the regime of 
our problem, we have no choice but to seek an alternative formulation. 
What is required is a theoretical framework that has the ability 
to sensibly and efficiently deal with multiple energy scales, 
as the perturbative approach does, 
but one that also does not rely \emph{ab initio} 
on the existence of a small coupling constant. 
The framework which possesses both of these characteristics is 
EFT, as we shall see.

\section{Effective Field Theory}
\label{EFT}

EFT is a technique for developing theories of problems with 
multiple energy scales. 
It is applicable in situations where we wish to understand 
the physics at some low-energy scale as the limiting case of a 
more general problem whose full features are apparent only at 
some higher energy.

For simplicity, consider a generic problem with two energy scales. 
The complete physics of the problem throughout the full spectrum of energies 
can be said to be described by some Lagrangian density 
$\mathcal{L}\,(\varphi)$ in terms of some degrees of freedom $\varphi$. 
At a characteristic energy scale $E_{under}$, 
the full properties of $\mathcal{L}$ are vital to an understanding of 
the physics. 
In general, we may or may not know $\mathcal{L}$. 
Even if we know $\mathcal{L}$, we may or may not be able to solve it 
for the dynamics of the underlying theory.
In our problem of the hadronic phase, for example, $\mathcal{L}$ is the 
QCD Lagrangian, and the underlying, full physics would be QCD. 
Processes explicitly involving quarks and gluons are seen to be important 
at energies of order $E_{under}\approx M_{QCD}$. 
At significantly lower energies, however, these processes ``freeze out'' 
and only hadronic degrees of freedom are available to the system. 
In the EFT, we therefore set some energy cutoff $\Lambda$ which divides 
energies of order $E_{under}$ from energies at which some of the freedoms 
of the full Lagrangian can be neglected. 

The $S$-matrix contains all of the information relating initial states 
to final states in a many-body problem. 
The elements of the matrix can be calculated from the path integral
\begin{equation}
Z=\int\mathcal{D}\varphi e^{i\int\mathrm{d}^D x\mathcal{L}(\varphi)},
\end{equation}
where $D$ is the number of spacetime dimensions in the problem. 
This expression is based in the full physics of the elementary, 
microscopic theory contained in $\mathcal{L}$.

We can begin to exploit the differences in physics between 
the two energy scales by using $\Lambda$ to relabel the fields 
according to momentum,
\begin{equation}
\varphi=\left\{ \begin{array}{ll}
\varphi_H & (p>\Lambda)\\
\varphi_L & (p<\Lambda),
\end{array}\right.
\label{decomp}
\end{equation}
and integrating over the degrees of freedom in $\varphi_H$. This gives us
\begin{equation}
Z=\int\mathcal{D}\varphi_L\int\mathcal{D}\varphi_H 
  e^{i\int\mathrm{d}^D x\mathcal{L}(\varphi_L,\varphi_H)} 
 = \int\mathcal{D}\varphi_L 
   e^{i\int\mathrm{d}^D x \mathcal{L}_{eff}(\varphi_L)},
\end{equation}
where $\mathcal{L}_{eff}$ is defined by
\begin{equation} \label{Leff}
\int\mathrm{d}^D x\,\mathcal{L}_{eff}(\varphi_L) 
= -i\ln\int\mathcal{D}\varphi_H 
   e^{i\int\mathrm{d}^D x\mathcal{L}(\varphi_L,\varphi_H)}.
\end{equation}
$\mathcal{L}_{eff}$ is obviously a function only of $\varphi_L$, 
since we have integrated over $\varphi_H$. 
We can create a series representation for Eq. (\ref{Leff}),
\begin{equation} \label{series}
\int\mathrm{d}^D x\,\mathcal{L}_{eff}(\varphi_L)
    \equiv\int\mathrm{d}^D x\sum_i g_i O_i(\varphi_L).
\end{equation}
In general, Eq. (\ref{series}) is a pure mathematical statement 
without any real physical content, but it becomes more meaningful 
when we make some statements about the properties of 
the operators $O_i$
and coefficients $g_i$.

The operators $O_i$ can in general be quite complicated. 
We can see, however, that they must possess two important physical properties. 
The first is that, although they contain an arbitrary number of derivatives, 
the $O_i$'s are in fact local operators in the sense that 
they involve fields at the same spacetime point. 
Nevertheless, their necessary dependence on derivatives  
follows from the fact that they depend only on fields with 
momentum components below $\Lambda$. 
As such, the uncertainty principle tells us that the operators can probe 
length scales only down to a minimum distance $\sim 1/\Lambda$, 
and so high-order derivatives will be necessary to perform the 
averaging over smaller length scales.
The idea is the same as that of a multipole expansion in 
electrodynamics.
The second observation we can make regarding the $O_i$'s is that, 
for an appropriate decomposition (\ref{decomp}), they must possess all 
of the symmetries and transformation properties of the underlying 
high-energy theory. 
Even if a particular symmetry is broken, it will manifest itself
in the same way in the effective Lagrangian.

The coefficients $g_i$ can likewise be put in perspective with 
some physical arguments. 
It is clear that they reflect the ``freeze out'' of the high-energy 
degrees of freedom below the cutoff $\Lambda$. 
Therefore, it is true that their particular form is dependent on 
the nature of the underlying theory and the structure of 
$\mathcal{L}\,(\varphi_L,\varphi_H)$. 
Furthermore, they are obviously a function of the cutoff parameter $\Lambda$. 
Note, however, that the specific dependence of the $g_i$ on $\Lambda$ 
is constrained by the familiar principle of renormalization-group 
invariance: physical observables cannot 
depend on the cutoff because the choice of the latter is arbitrary.

The physical motivation for EFT should now be apparent. 
It is an extremely common circumstance in physics that we wish to treat 
some low-energy limit of a fundamentally higher-energy problem. 
In such a case, we expect that the low-energy physics will prove to be 
some reflection of the full problem, but with the phase space 
restricted such that high-energy excitations or degrees of freedom 
are not accessible. 
Indeed, this is arguably the case in any physical problem short of a 
theory of everything: there exists some smaller-length and higher-energy scale 
at which new physics occurs, but whose details are unimportant to a 
description of a particular system. 
Were this not the case, physics as we know it would be impossible.
The power of EFT, then, is that not only does it provide a framework to 
unify a problem with the energy scale above it, but it turns the energy 
difference from an obstacle into part of the solution: 
it recognizes this principle of limiting the system's freedom 
at low energy and incorporates it directly into the theoretical strategy.

The EFT is useful even when we know and can solve the underlying theory at 
higher energies. 
In this case, it is possible to derive the coefficients $g_i$ 
from our knowledge of the underlying Lagrangian $\mathcal{L}\,(\varphi)$. 
The advantage lies in the reorganization of the dynamics in the EFT,
as it is often the case that the effective degrees of freedom
are ``collective'' excitations of the underlying degrees of freedom.
This is the case whenever there is spontaneous breaking
of a continuous symmetry and Goldstone bosons appear.
An example is magnons, which are
a better way to describe low-energy excitations in a system
with spontaneous magnetization than individual magnetic moments.
Another example are pions in QCD.

But EFT is a necessity when either the underlying theory is not known or, 
as in the case of QCD, it is not currently solvable. 
In this case, we can rely on Weinberg's ``theorem'':
``if one writes down the most general possible Lagrangian, 
including \emph{all} terms consistent with assumed symmetry principles, 
and then calculates $S$-matrix elements with this Lagrangian to any order 
in perturbation theory, the result will simply be the most general possible 
$S$-matrix consistent with analyticity, perturbative unitarity, 
cluster decomposition and the assumed symmetry principles'' \cite{weinberg79}. 
There is no known general proof of this ``theorem'', although it has been 
proven for a scalar field with $Z_2$ symmetry in Euclidean space \cite{ball94}.
Nevertheless, it is certainly plausible, as it is really stating that the 
most general quantum field theory is a direct consequence only of 
analyticity, unitarity, cluster decomposition, and symmetry. 
This, combined with the fact that there are no known counterexamples, 
suggests that we would do well to consider the ramifications of the 
``theorem'' regardless of its current lack of rigor.

Weinberg's ``theorem'' allows us to formulate a plan to solve problems 
with an unknown or insoluble Lagrangian by using EFT.
\begin{enumerate}
\item Identify the relevant degrees of freedom and symmetries of the problem.
\item Construct the most general Lagrangian consistent with these limitations.
\item Do standard quantum field theory with this Lagrangian.
\end{enumerate}
``Standard quantum field theory'' consists of computing the contributions 
from all diagrams with momenta $Q<\Lambda$, and then renormalizing the 
result to relate the coefficients $g_i$ to the physical observables of 
the problem. 
After renormalization, observables should be independent of $\Lambda$. 
According to Weinberg's ``theorem'', these steps give the dynamics of the 
general system, and the process of building an EFT will have introduced 
no spurious information.

An important issue is that of ordering the
infinite number of contributions to any observable, 
known as ``power counting''.
The minimal assumption is 
that, barring a suppression by some symmetry, the coefficients in the 
expansion will be roughly of ${\cal O}(1)$ once expressed
in units of $E_{under}$ according to dimensional analysis. 
This is called the assumption of ``naturalness'', and its validity 
is a consequence of our having chosen an appropriate cutoff $\Lambda$. 
The perturbative result of this process is a controlled expansion 
in energy, $E/E_{under}$.
However, since observables are cutoff independent,
the same expansion is valid for any cutoff.
As we vary the cutoff, strength moves from one contribution
to another that appears at the same order in $E/E_{under}$.
So, regardless of the cutoff, to any given order in $E/E_{under}$
only a finite number of $g_i$'s need to be considered.
One can use a finite number of experimental data as input
to determine these $g_i$'s, and then 
use the known $g_i$'s to predict everything else,
with an error given by the estimated size of higher-order terms.
Thus, the most important ingredient to the final stage of building the 
EFT is power counting.

The controlled, natural expansion resulting from an 
EFT can be contrasted favorably with the characteristics of 
a theory constructed with traditional model-building strategies. 
A successful model takes a complex problem and reduces it reasonably to only 
a few degrees of freedom and types of interactions. 
This is a valid way to build a physical understanding of the problem, 
in that it allows the identification of the important symmetries 
and degrees of freedom. 
It falls short in quantitative predictive power, however, because 
there is no way to place a bound on the error arising 
from effects not included in the model. 
EFT, on the other hand, provides a controlled expansion of the
most general dynamics, 
so errors are well bounded and predictable. 
The value of such an advantage in comparing theoretical 
predictions to experiment cannot be overstated.

\subsection{A classical example}

At this point, some simple examples of EFT calculations will prove fruitful. 
We begin with a problem drawn from classical physics, 
a light object interacting via gravity with a much larger object, 
for example an apple near the surface of the earth. 
We can easily identify the relevant degree of freedom and symmetries 
of this simple problem. 
By experimenting with various objects close
to the surface of the earth, we find that, 
to a good approximation,
the degree of freedom is the mass $m$, 
while the symmetries involve translations parallel to the earth's surface 
and rotations about an axis normal to it. 
According to our recipe for an effective theory,
we write down the most general potential 
reflecting these properties which 
is a power series in the height $h$: 
\begin{equation}
V_{eff}(h)=m\sum_{i=0}^\infty g_i h^i.
\label{effgravpot}
\end{equation}
Each of the $g_i$'s are parameters that can be fit to experimental data.
(Here we are of course neglecting any quantum corrections,
so there are no terms which are non-analytic in $h$
and $V_{eff}$ is directly observable.)
The first term is an irrelevant constant that depends on the arbitrary
choice of the zero in energy.
The second term is the familiar $mgh$, where $g$ is the acceleration of a 
free body due to gravity as measured at $h=0$.
The linear form of the gravitational potential, however, is simply 
an approximation: higher-order terms
are corrections that could in principle be extracted from careful
measurements.
This expansion will breakdown at some
energy $E_{under}$, which can be used to define
a distance scale $R$ through $E_{under}\equiv mgR$.
This is a controlled expansion in $h$ 
suitable for the low-energy regime $E\ll E_{under}$. 
The assumption of naturalness means
\begin{equation}
\frac{m g_{i+1} h^{i+1}}{m g_i h^i}=c_i\frac{E}{E_{under}}=c_i\frac{h}{R},
\end{equation}
where the $c_i$'s are coefficients of ${\cal O}(1)$. 
This reduces to 
\begin{equation}
g_{i+1}= {\cal O} \left( \frac{g}{R^{i}}\right).
\label{gravnat}
\end{equation}

Thanks to Newton's consideration of large distances, 
in this case we know the underlying theory. 
A more accurate expression for the gravitational potential 
follows from Newton's law of universal gravitation,
\begin{equation} \label{newton}
V(h)=-G M m \frac{1}{R+h},
\end{equation}
where $G$ is the gravitational constant, $M$ is the mass of the earth, 
and $R$ is its radius. 
For small $h$, we can expand Eq. (\ref{newton}) as
\begin{equation} \label{undergrav}
V(h)=mgR \sum_{i=0}^\infty (-1)^{i-1} \left(\frac{h}{R}\right)^i,
\end{equation}
where we have simplified the notation with the relation 
$g\equiv G M/R^2$. 
The underlying potential (\ref{undergrav}) indeed 
gives the effective gravitational potential (\ref{effgravpot})
for $h\ll R$. 
Matching the two expressions,
the coefficients of the effective potential
are easily seen to be
\begin{equation}
g_{i+1}= (-1)^{i} \frac{g}{R^{i}}.
\end{equation}
The naturalness assumption 
(\ref{gravnat}) is verified ---all of the $c_i$'s are $\pm 1$---
with the added understanding of the scale $R$ 
as the earth's radius.

We have seen that the familiar linear form of gravitational potential energy 
can be treated as an effective theory of the more general theory given by 
Newton's law of gravitation. 
It is worth noting in passing, however, that Newton's law could 
also be viewed 
as an effective theory of a general relativistic formulation of gravitation 
whose effects are only visible at even larger energies. 
Thus, we can begin to develop a picture of nature as an ``onion'', 
with each successively smaller energy scale being 
described by an effective theory of the last.

\subsection{Non-relativistic QED}

While a simple example from classical mechanics is instructive, we are
now ready to increase our insight into the nature of EFT by examining a 
quantum-mechanical problem that bring us closer to our 
nuclear-physics interest.
Let us attempt to find the effective Lagrangian
for QED at low energies.

Consider fermions described by a field $\psi$
of mass $m$ and charge $e$, and assume 
the underlying theory to be given by the QED Lagrangian
\begin{equation}
\mathcal{L}_{QED}=\overline{\psi}[i(\not{\!\partial}-ie\not{\!\!A})-m]\psi
+\frac{1}{4}F_{\mu\nu}F^{\mu\nu},
\end{equation}
where $A_\mu$ ($F_{\mu\nu}$) is the photon field (field strength).

\begin{figure}[tbp]
\includegraphics{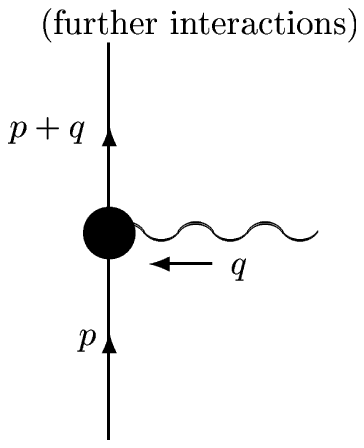}
\caption{Schematic representation of a class of processes
involving a low-momentum fermion. 
The fermion first interacts with a real or virtual photon
and then propagates until it interacts further.}
\label{qedfig} 
\end{figure}

What happens in processes where all particles have momenta of the same order
$Q\ll m$?
Consider the part of a
diagram found in Fig. \ref{qedfig},
where the fermion interacts with a low-energy photon,
either real or virtual. 
In this non-relativistic regime we can write 
\begin{eqnarray}
&&|\vec{p}| \sim |\vec{q}| \sim Q\ll m,\\
&&p^0=\sqrt{\vec{p}^2+m^2}\sim m+ {\cal O}\left(\frac{Q^2}{m}\right),\\
&&q^0 \sim |\vec{q}| \sim Q.
\end{eqnarray}
The internal fermion line (with momentum $p+q$) contributes 
a factor to these diagrams
\begin{eqnarray}
\frac{i}{\not{\!\!p}+\not{\!q}-m+i\epsilon}
&=&\frac{i(\not{\!\!p}+\not{\!q}+m)}{(p^0+q^0)^2
        -(\vec{p}+\vec{q})^2-m^2+i\epsilon}\nonumber\\
&=&\frac{i(\not{\!\!p}+\not{\!q}+m)}{{p^0}^2+2q^0p^0+{q^0}^2
          -(\vec{p}+\vec{q})^2-m^2+i\epsilon}\nonumber\\
&=&\frac{i(\not{\!\!p}+m)}{2p^0q^0+i\epsilon}+\ldots
=\frac{i}{q^0+i\epsilon}\frac{1+\gamma^0}{2}+\ldots
\label{internalqed}
\end{eqnarray}
Here ``$\ldots$'' stand for terms which are suppressed by powers of $Q/m$.

The last equality in Eq. (\ref{internalqed}) contains the operator 
we will call $P_+$, which can be defined together with the complementary $P_-$,
\begin{equation}
P_\pm\equiv\frac{1\pm\gamma^0}{2}.
\end{equation}
It is easy to verify that these operators possess the properties of 
idempotent projection operators:
\begin{equation}
P_\pm^2=P_\pm, \;\;
P_\pm P_\mp=0.
\end{equation}

The presence of $P_+$ in Eq. (\ref{internalqed}) indicates that
the fermion propagation is given by 
the particle propagating forward in time.
This suggests that we separate the nearly-inert mass $m$ and
the antiparticle component from the field, by
rewriting the $\psi$ in 
a ``heavy-fermion formalism'' \cite{georgi}
\begin{equation}
e^{-i m t}\smash[b]{\underset{\sim}{\smash[b]\psi}}_\pm\equiv P_\pm\psi.
\end{equation}
The Lagrangian then becomes
\begin{equation}
\mathcal{L}_{QED}=
\overline{\smash[b]{\underset{\sim}{\smash[b]\psi}}}_+
i\partial_0\smash[b]{\underset{\sim}{\smash[b]\psi}}_+ 
-\overline{\smash[b]{\underset{\sim}{\smash[b]\psi}}}_-
i\vec{\gamma}\cdot\vec{\nabla}\smash[b]{\underset{\sim}{\smash[b]\psi}}_+ 
+(\overline{i\vec{\gamma}\cdot
\vec{\nabla}\smash[b]{\underset{\sim}{\smash[b]\psi}}_-})
\smash[b]{\underset{\sim}{\smash[b]\psi}}_+ 
+\overline{\smash[b]{\underset{\sim}{\smash[b]\psi}}}_-
(-i\partial_0-2m)\smash[b]{\underset{\sim}{\smash[b]\psi}}_-+\ldots
\end{equation}
We calculate the low-energy Lagrangian from 
Eq. (\ref{Leff}). 
Completing the square, doing the gaussian integral over 
$\smash[b]{\underset{\sim}{\smash[b]\psi}}_-$, and
renaming 
$\smash[b]{\underset{\sim}{\smash[b]\psi}}_+$ to $\psi$
yields
\begin{equation} \label{qedleff}
\mathcal{L}_{eff}\,(\psi)
=\overline{\psi}i(\partial_0-ieA_0)\psi
+\frac{1}{2m}\overline{\psi}(\vec{\nabla}-ie\vec{A})^2\psi+\ldots
+ \frac{1}{4}F_{\mu\nu}F^{\mu\nu} +\ldots
\end{equation}

Of course, in the real world,
there exist more than one particle
coupling to photons, and other gauge bosons coupling
to these particles.
Because the photon is massless while weak bosons are not,
the latter can be integrated out at low energies.
Likewise, we can integrate out other, heavier fermions.
This results in additional terms in the effective Lagrangian,
such as a Pauli term that gives rise to an anomalous magnetic moment
through a parameter $\kappa$:
\begin{equation}
\mathcal{L}_{eff}^{(\mathrm{higher})}\,(\psi)
=\frac{e}{2m}\kappa\varepsilon_{ijk}\overline{\psi}\sigma_k\psi F_{ij}
+\ldots \label{qedleffhigh}
\end{equation}
The new parameters  
can be calculated from the underlying theory.
All interactions in $\mathcal{L}_{eff}^{(\mathrm{higher})}$
are suppressed by powers of a heavy mass.
The Pauli term, for example, is $\propto 1/m$, and $\kappa$
is of ${\cal O}(1)$  unless the particle represented 
by $\psi$ has only weak interactions.

Now, according to Weinberg's ``theorem'', we do not need to go through
this whole song and dance.
If we directly construct  
the most general effective Lagrangian involving the fermion $\psi$ 
and photon $A$ that is invariant under gauge 
transformations, parity, time reversal, and (non-relativistic) Lorentz boosts,
we find the $\mathcal{L}_{eff}$ above. 

\subsection{Bound states in QED}

With the above effective Lagrangian,
we can return to the problem of the hydrogen atom,
considered in the Introduction.
We have already seen that this problem possesses multiple energy scales, 
much like the nuclear-physics problem we wish to solve. 
Discussing non-relativistic bound states in QED
is a good warm-up for tackling nuclear bound states.

\begin{figure}
\includegraphics{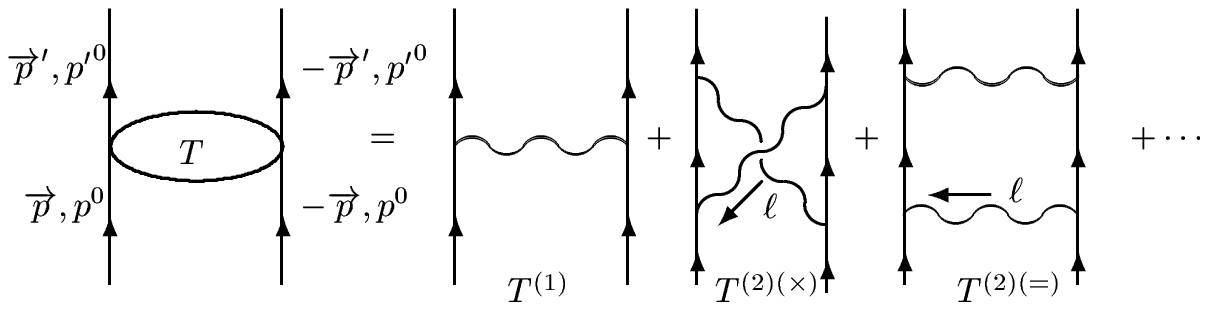}
\caption{Diagrams representing all photon-exchange interactions between 
two heavy fermions in the center of momentum frame. 
The interaction $T$ is shorthand for all interactions which can occur. 
Note that $\ell$ is a dummy 4-momentum, and as such should be 
integrated over in the two loop diagrams.}
\label{boundfig}
\end{figure}

For simplicity, let us consider the interaction of
two non-relativistic fermions of the same type.
We denote the
initial (final) center-of-mass momentum $p$ ($p'$),
with
$|\vec{p}|\sim|\vec{p}'|\sim Q$ and 
$p^0\sim {p'}^0\sim Q^2/2m$.
Since heavy-fermion number is conserved,
all diagrams that contribute 
to the scattering amplitude $T$ have two fermion lines 
that go through.
Let us consider first the diagrams made of photon exchange only,
that is, of the type in Fig. \ref{boundfig},
where the photon-fermion vertex
comes from the first term in the Lagrangian (\ref{qedleff}).
We want to estimate each to leading order in $Q$. 

The first-order diagram contributes a term
\begin{eqnarray}
T^{(1)}
&=&\frac{e^2}{(p-p')^2+i\varepsilon}
=\frac{e^2}{(p^0-{p'}^0)^2-(\vec{p}-\vec{p}')^2+i\varepsilon}
= -\frac{e^2}{(\vec{p}-\vec{p}')^2-i\varepsilon} +\ldots \nonumber \\
&&\sim  \frac{e^2}{Q^2}. 
\label{T1}
\end{eqnarray}

The second-order term in which the photon lines cross is
\begin{multline} \label{T2x}
T^{(2)(\times)}
= -i e^4\int\frac{\mathrm{d}^4 \ell}{(2\pi)^4}
  \frac{1}{p^0+\ell^0-\frac{(\vec{\ell}+\vec{p})^2}{2m}+i\varepsilon} 
  \times
  \frac{1}{p'^0+\ell^0-\frac{(\vec{\ell}-\vec{p}')^2}{2m}+i\varepsilon}\\ 
  \times
  \frac{1}{(p^0-{p'}^0+\ell^0)^2-(\vec{p}-\vec{p}'+\vec{\ell})^2+i\varepsilon} 
  \times
  \frac{1}{{\ell^0}^2-{\vec{\ell}}^2+i\varepsilon}.
\end{multline}
The integral in $\ell^0$ can be evaluated as usual with a contour integral. 
It is clear that the first and second factors possess poles below 
the real axis, while the third and fourth terms each produce a pole 
above and below the real axis. 
We therefore minimize the algebra by closing in an infinite semicircle 
above the real axis: this way we avoid the two shallow poles from the nucleon
propagators, and
\begin{gather}
\begin{split}
T^{(2)(\times)} = 
e^4\int\frac{\mathrm{d}^3\ell}{(2\pi)^3} 
\frac{1}{p^0-|\vec{\ell}|-\frac{(\vec{\ell}+\vec{p})^2}{2m}+i\varepsilon} 
\times
\frac{1}{p'^0-|\vec{\ell}|-\frac{(\vec{\ell}-\vec{p}')^2}{2m}+i\varepsilon}
\\
\times
\frac{1}{(p^0-{p'}^0-|\vec{\ell}|)^2-(\vec{p}-\vec{p}'+\vec{\ell})^2
+i\varepsilon} 
\times
\frac{1}{2|\vec{\ell}|-i\varepsilon}
+ \ldots\\
\end{split} \nonumber \\
\sim e^4\Big(\frac{Q^3}{4\pi}\Big)
\Big(\frac{1}{Q}\Big)\Big(\frac{1}{Q}\Big)
\Big(\frac{1}{Q^2}\Big)\Big(\frac{1}{Q}\Big)
\sim\alpha\frac{e^2}{Q^2}.
\label{T2xint}
\end{gather}
There is no surprise here: 
these contributions are, as ordinarily expected from loop diagrams, down
by a factor of $\alpha$ compared to Eq. (\ref{T1}).

Life is not so boring, fortunately.
In the other second-order term, the photon lines do not cross. 
This contributes 
\begin{multline} \label{T2=}
T^{(2)(=)}= 
-ie^4\int\frac{\mathrm{d}^4\ell}{(2\pi)^4}\frac{1}{p^0+\ell^0
-\frac{(\vec{\ell}+\vec{p})^2}{2m}+i\varepsilon} 
\times\frac{1}{p^0-\ell^0-\frac{(\vec{\ell}+\vec{p})^2}{2m}+i\varepsilon}\\
\times\frac{1}{(p^0-{p'}^0+\ell^0)^2-(\vec{p}-\vec{p}'+\vec{\ell})^2
               +i\varepsilon} 
\times\frac{1}{{\ell^0}^2-{\vec{\ell}}^2+i\varepsilon}.
\end{multline}
The pole structure of Eq. (\ref{T2=}) is the same as that of Eq. (\ref{T2x}), 
except that the pole of the second factor lies in the upper half complex plane.
In addition to residues from poles from photon propagators
that contribute terms similar to the ones in Eq. (\ref{T2xint}),
we cannot avoid a contribution from a shallow pole,
\begin{gather}
\begin{split}
T^{(2)(=)}= 
e^4\int\frac{\mathrm{d}^3\ell}{(2\pi)^3}
\frac{1}{2p^0-\frac{(\vec{\ell}+\vec{p})^2}{m}+i\varepsilon}
\times\frac{1}{\Big[2p^0-{p'}^0-\frac{(\vec{\ell}+\vec{p})^2}{2m}\Big]^2 
               -(\vec{p}-\vec{p}'+\vec{\ell})^2+i\varepsilon}\\ 
\times\frac{1}{\Big[p^0-\frac{(\vec{p}+\vec{\ell})^2}{2m}\Big]^2
                -{\vec{\ell}}^2+i\varepsilon}
+\ldots
\end{split} \nonumber\\
\sim e^4\Big(\frac{Q^3}{4\pi}\Big)\Big(\frac{m}{Q^2}\Big)
        \Big(\frac{1}{Q^2}\Big)\Big(\frac{1}{Q^2}\Big)+\alpha\frac{e^2}{Q^2}
\sim\alpha\frac{e^2}{Q^2}\Big(\frac{m}{Q}+ \ldots\Big).
\end{gather}

We note that
\begin{equation}
\frac{m}{Q}\gg 1.
\end{equation}
Thus, the term in $T^{(2)(=)}$ from the shallow pole is enhanced over 
the other terms. 
Having done the time-component integrals in these diagrams, we can now think 
in terms of \emph{time-ordered} perturbation theory. 
The reason time-ordered diagrams are useful here is that we
are considering the interactions of non-relativistic particles.
In time-ordered perturbation theory loops correspond
to three-dimensional integrals.
Each diagram in covariant perturbation theory unfolds into
various time-ordered diagrams.
The time-ordering of the four vertices in each second-order diagram is 
significant. 
Second-order terms, then, can be said to fall into two groups, 
those for which both photons are created and then both are destroyed, 
and those for which the first photon is destroyed before the second is created.
These second diagrams are really iterations of first-order diagrams, 
and it is these that lead to the enhancement. 
It is clear that they can only occur as part of $T^{(2)(=)}$, 
since $T^{(2)(\times)}$ diagrams contain a crossing of photon lines 
by definition.

Infrared enhancements will appear at all orders
in perturbation theory, always from ``two-fermion reducible'' diagrams.
The leading terms at each order form a series that goes roughly as 
\begin{equation} \label{T}
T
\sim\frac{e^2}{Q^2}\left[1+ {\cal O}\left(\alpha\frac{m}{Q}\right)
+\cdots+{\cal O}(\alpha) \right]
\sim\frac{e^2}{Q^2}\left[\frac{1}{1-{\cal O}\left(\alpha\frac{m}{Q}\right)}
+\cdots+{\cal O}(\alpha) \right].
\end{equation}
It is clear from Eq. (\ref{T}) that the series expansion
breaks down at sufficiently small momenta such that
$m/Q$ compensates for $\alpha$, that is, at
\begin{equation}
Q\sim\alpha m,
\end{equation}
indicating the existence of
a bound state with binding energy
\begin{equation}
B\sim\frac{Q^2}{m}\sim\alpha^2 m.
\end{equation}
Hence we reproduce the estimates (\ref{hbound}) and (\ref{hboundbound})
arrived at earlier.

\begin{figure}[t]
\includegraphics[totalheight=1.5in]{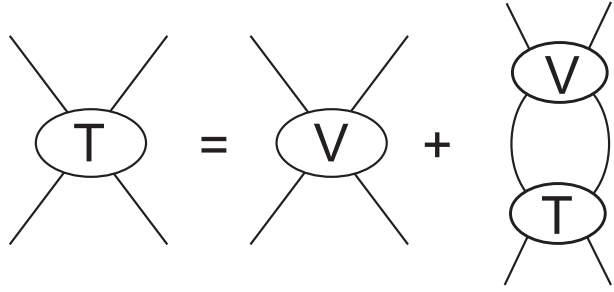}
\caption{The Lippmann-Schwinger equation with potential $V$.
\label{potential}}
\end{figure}

\begin{figure}[t]
\includegraphics[totalheight=1.0in]{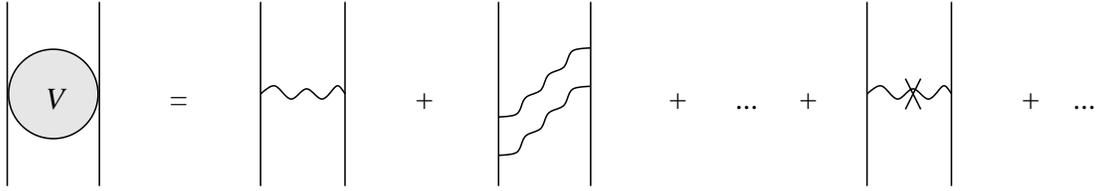}
\caption{Some time-ordered diagrams contributing to the two-particle potential
$V$ in non-relativistic QED.
All orderings with at least one photon
in intermediate states are included.
A cross denotes energy corrections.
\label{qedpot}}
\end{figure}

This is {\it not} just a more complicated way of generating
known results: it is a more {\it fundamental} way.
Of course, the actual resummation of this series can only be carried out
by explicitly solving the Lippmann-Schwinger (or, equivalently, 
the Schr\"odinger) equation with a potential $V$,
as represented in Fig. \ref{potential}.
The potential is given in leading order by 
the Coulomb potential given in Eq. (\ref{T1}).
But from quantum mechanics alone we are at a loss about how to
improve on the Coulomb potential.
With non-relativistic QED, the prescription is clear,
and it is illustrated in Fig. \ref{qedpot}.
The Coulomb potential of ${\cal O}(e^2/Q^2)$ is the result
from the lowest-order, static photon exchange.
The corrections in the potential come from all other two-fermion irreducible
time-ordered diagrams.
As we have seem above,
the irreducible second-order diagrams generate
corrections
of ${\cal O}(\alpha e^2/Q^2)$.
The recoil part of the one-photon exchange (\ref{T1})
is an ${\cal O}(Q/m \times e^2/Q^2)$ correction,
which for the bound state (where $Q/m \sim \alpha$)
is comparable to two-photon exchange.
And so on.

A remarkable feature of the discussion above is that 
all the photon-exchange diagrams considered scaled with
negative powers of $Q$. This is rooted on our 
temporary restriction to diagrams built from
the simplest interaction in the Lagrangian (\ref{qedleff}).
It means that no new ultraviolet cutoff 
is needed in loop integrals, once the bare fermion charge and mass
are adjusted to remove cutoff dependence in diagrams
involving a single fermion.
At some point in the past great importance was attributed to
this ``renormalizability''. 
Nowadays, we realize that its simple form in QED is 
limited.
Interactions with more derivatives, which
are not forbidden by symmetry and thus exist, spoil
this simplification. 
Piling up sufficiently many higher-derivative interactions from 
Eq. (\ref{qedleffhigh}),
a two-fermion diagram will scale as a positive power 
of $Q$, and diverge in the absence of a cutoff.
However, because four-fermion interactions
can be constructed that are gauge invariant, the new
cutoff dependence can still be eliminated by a shift in
four-fermion parameters. 
The EFT is still renormalizable, in the sense that at any
given power of $Q$ a finite number of parameters
will remove cutoff dependence from observables
(renormalization-group invariance).
QED can be deceivingly simple because the symmetries
allow couplings of dimension four, which dominate.
As we are going to see next, the same does {\it not} happen
in the EFT of QCD.

\section{Building a QCD-based EFT}

\subsection{Standard model and QCD}

We now focus on the energy region of interest. 
The EFT corresponding to the standard model at 
an energy scale of a few GeV involves only the lightest
leptons and quarks, gluons, and the photon; 
weak gauge bosons, and heavy leptons and quarks can be
integrated out.  
For simplicity, here we focus on interactions involving
the lightest $u$ and $d$ quarks. 
They can be arranged in
a flavor doublet $q=(^u_d)$. 
The relevant pieces of the effective Lagrangian at this scale are 
\begin{eqnarray}
 {\cal L} & = & 
    -\sum_{f=1}^3 \bar{l_f}(\partial\!\!\!\!\!/ -ie A \!\!\!\!/+ m_f )l_f 
    -\frac{1}{4}F_{\mu\nu}F^{\mu\nu}
    \nonumber  \\ 
& & -\bar{q}(\partial\!\!\!\!\!/-ig_{s}G\!\!\!\!\!/-ieQA\!\!\!\!/ )q 
    -\frac{1}{2}(m_{u}+m_{d})\bar{q}q 
    +\frac{1}{2}(m_{d}-m_{u})\bar{q}\tau_{3}q  \nonumber  \\ 
& & -\frac{1}{2}Tr[G_{\mu\nu}G^{\mu\nu}]
    +\frac{\bar{\theta} g_{s}^{2}}{32 \pi^{2}}\varepsilon_{\mu\nu\rho\sigma}
    Tr[G^{\mu\nu}G^{\rho\sigma}] 
    +\ldots 
\label{intro9}
\end{eqnarray}
\noindent
Here the $l_f$'s are the lepton fields with mass $m_f$,
$G_{\mu}$ ($A_{\mu}$) is the gluon (photon) field of strength $G_{\mu\nu}$ 
($F_{\mu\nu}$),
$\tau_3$ is the usual Pauli matrix,
$Q=1/6+\tau_{3}/2$  is the quark charge matrix,
and ``$\ldots$'' denote higher-dimension terms.   
Higher-dimension interactions are suppressed by
powers of the masses of the heavy particles that have been integrated out.
We will neglect them in a (very good) first approximation.
We will also neglect the theta term, since the strong
CP parameter $\bar{\theta}$ is found to be unnaturally small 
(the so-called strong CP problem).
The effect of these terms can be easily incorporated in a more
comprehensive analysis.

The leading Lagrangian (QCD + QED) becomes then invariant
under parity and time-reversal transformations.
The quark/gluon sector has
4 remaining parameters: the gauge 
couplings for strong ($g_{s}$) and electromagnetic ($e$) interactions,
and the masses of the up ($m_{u}$) and the down ($m_{d}$) quarks. 

In the chiral limit, that is, when  $m_{u} = m_{d} = 0$ and $e=0$, 
the action of the QCD Lagrangian becomes invariant under 
scale transformations: 
$x \rightarrow \lambda^{-1} x$, $q \rightarrow \lambda^{3/2}
q$, and $G_\mu \rightarrow \lambda G_\mu$. 
Thus there is no dimensionful
parameter in the Lagrangian; there is only one dimensionless parameter
which is the strong-interaction coupling constant $g_s$. 
However, the
scale symmetry is explicitly broken due to quantum corrections
that probe high energies. 
A regulator is
necessary, thus introducing a dimensionful parameter in the problem. 
As a
result, $g_s$ becomes a ``running coupling'' that
is a function of a momentum scale. 
We observe in fact that $g_s$ grows as the scale is
lowered, and reaches $g_s \sim 1$ near 1 GeV.
Perturbation theory in $g_s$, a useful tool at high energies,
fails miserably at nuclear energies.

\subsection{Basic assumptions about a QCD-based EFT}

Thus in our pursuit to build an EFT, we have to take the properties of
the nonperturbative QCD into account,
even though we cannot at present deduce them. 
The first
assumption we make is that QCD ``confines'': only
colorless states (hadrons) are 
asymptotic states.  
In the low-energy EFT, only hadrons need to be incorporated as fields.
Our second assumption is that QCD is natural.
Since almost all
hadron masses $\gaprox1$ GeV, we
conclude that there is a characteristic
QCD mass scale $M_{QCD} \sim 1$ GeV. 
This sets the limit of validity of the EFT.

Because we want to discuss nuclei, we certainly want
to include the nucleon in the EFT. 
Because the delta-nucleon mass difference
$\delta \equiv m_\Delta -m_N \simeq 300$ MeV $\ll M_{QCD}$,
if we want to explore the full region of energies accessible to EFT,
we incorporate the delta isobar as well.
Nucleon and delta fields
form isospin multiplets as following:
\begin{eqnarray}
N = \bigl(\begin{smallmatrix}p\\n\end{smallmatrix}\bigr) &
\;\;\;\;\;\;\;\; {\rm and} \;\;\;\;\;\;\;\;\;\;\;\Delta =
\L(\begin{smallmatrix}\Delta^{++}\\\Delta^{+}\\ \Delta^{0}\\
\Delta^{-} \end{smallmatrix}\R).
\end{eqnarray}
Below we denote these fermions by $\psi$.
Other baryon states such as $N^*$ 
are not included, as their 
mass differences to the nucleon approach $M_{QCD}$
($m_{N^*}-m_N \simeq 500$ MeV, etc.).
Mesons such as the $\rho$ and the $\omega$
are not included either because they
have masses ${\cal O}(M_{QCD})$.
(Moreover, they interact
with the pions and nucleons via dimension-four interactions that are not
weak, which is at present an insurmountable obstacle for a
systematic approach.) 

Does the pion mass, 
$m_\pi \simeq 140$ MeV $\ll M_{QCD}$, imply a breakdown of
naturalness? No!
We notice that in the chiral limit the leading-order Lagrangian
(\ref{intro9}) has an
$SU_L(2)\times SU_R(2) \sim SO(4)$ chiral symmetry,
since it is invariant under independent
$SU(2)$ rotations of left- and right-handed
quarks,
\begin{equation} \label{chsym}
q_{R(L)} \equiv \frac{1 + (-) \gamma_5}{2} q
\; \rightarrow \exp(i \vec{\alpha}_{R(L)}\cdot \vec{\tau}) \; q_{R(L)}.
\end{equation}
This symmetry is not manifest in the hadron
spectrum. There is, for example,
no scalar particle degenerate with the three pions.
We thus make our third assumption,
that chiral symmetry is spontaneously broken 
down to its diagonal subgroup,
the $SU_{L+R}(2)\sim SO(3)$ of isospin.
Although we cannot at present calculate it,
the effective potential of QCD, when plotted as function
of quark bilinears ($\bar{q} \boldtau i\gamma_5 q$, $\bar{q}q$),
has to have a ``Mexican-hat'' shape.
The symmetry of the potential is $SO(4)$.
The degenerate minima form thus a 
(four-dimensional) ``chiral circle'',
the radius of which we call $f_\pi$.
(Later, we find out that $f_\pi$ so defined coincides
with the pion decay constant.)
The symmetry is broken because
the world actually sits on one particular point
of this circle (which is selected by the quark-mass perturbation).
Excitations orthogonal to the circle
should have mass $m_\sigma \sim M_{QCD}$, since this is the scale that
sets the curvature of the effective potential.
Excitations along the circle, on the other hand,
are massless; the three Goldstone bosons can be associated with
the light pions.
 
Interactions of pions in the EFT have to reproduce these
symmetry properties.
There are several possible
parametrizations of the chiral circle
which correspond to different choices of pion fields $\boldpi$,
such as the $\sigma$-model-like, the
Callan-Coleman-Wess-Zumino construction, and the stereographic
projection. 
Since a chiral transformation rotates the chiral circle,
the infinitesimal transformation of the pions is
$\boldpi \rightarrow \boldpi +\boldepsilon$.
The EFT Lagrangian has to have the same symmetry,
which implies that the pion fields can be chosen
to couple derivatively.
This is very important
because it means that pion interactions involve the small momentum $Q$.
Moreover, because the derivative is on the chiral circle,
non-linear terms 
\begin{equation}
D^{-1}\equiv 1- \boldpi^2/4f_\pi^2 + \ldots
\end{equation}
appear.
There is a well-defined procedure to construct
the Lagrangian, the theory of non-linear realizations
of a symmetry \cite{CCWZ}.
We first define covariant objects, such as covariant derivatives
that involve the associated non-linear terms:
the pion covariant derivative ${\boldD}_\mu$ 
and the fermion covariant derivative ${\cal D}_\mu$.
Then, the invariant Lagrangian can be constructed
as the most general $SO(3)$-invariant Lagrangian 
made out of covariant objects.

We can now relax the restriction to the chiral limit
and consider the effect of non-vanishing quark masses.
The important point is that
the mass terms break $SO(4)$ in a specific way.
The common-mass term is the fourth component of an
$SO(4)$ vector, and the mass-difference term
is the third component of another $SO(4)$ vector.
The common-mass term, for example, tilts the Mexican
hat in the $\bar{q}q$ direction.
The chiral circle is no longer degenerate,
and the pions get a common mass $m_\pi^2 \sim (m_u + m_d) M_{QCD}$.
In the EFT, we construct all operators that break
chiral symmetry in the same way.
Their coefficients will be proportional
to powers of 
$(m_u + m_d)$
and $(m_d -m_u) \equiv \varepsilon (m_u + m_d)$. 

Finally, electromagnetic
interactions via ``soft'' photons
can be added as well through all gauge-invariant terms.
The one complication
is that the integrating out of ``hard'' photons
generates operators that do not necessarily involve
soft photons.
Such ``indirect'' electromagnetic operators 
can be constructed by looking at the 
chiral transformation properties of the (non-local) four-quark
operators they produce.
One can show that these operators break
chiral symmetry as 34 and 34-34 components of 
$SO(4)$ antisymmetric tensors.
The corresponding operators in the EFT
will break isospin with strengths proportional
to powers of $e^{2}$. 

To sum it up, 
the EFT Lagrangian has two classes of interactions.
One class is chiral invariant; 
they involve powers of the momentum and no dimension-four interactions are
permissible. 
Another class is chiral-symmetry breaking;
dimension-four operators are among the ones allowed,
but they are suppressed by powers of the small
quark masses or charges.
As a consequence,
all pion interactions 
are weak at low energies.

\subsection{The chiral Lagrangian}

Based on the above, the most general Lagrangian takes the 
schematic form:
\begin{eqnarray}
{\cal L} = 
\sum^\infty_{m n p q f}& & C_{m n p q f} 
\L(\frac{{\boldD}_\mu}{M_{QCD}} \R)^n 
\L(\frac{{\cal D}_\nu}{M_{QCD}} \R)^m 
\L(\frac{{\bar\psi} \psi}{f_\pi^2 M_{QCD}}\R)^\frac{f}{2} 
\L(\frac{\delta}{M_{QCD}}\R)^q
\L(\frac{m_\pi^2}{M_{QCD}^2}\frac{\boldpi^2}{f_\pi^2}\R)^p \nonumber \\
&& \times f_\pi^2 M_{QCD}^2,
\end{eqnarray}
where the $C_{m n p q f}$ are the unknown constants in the EFT. 
By naturalness these are of
${\cal O}(1)$ if isospin conserving
and ${\cal O}(\varepsilon)$ and ${\cal O}(e^2)$ if isospin breaking. 

It is convenient to introduce
the ``chiral index'' $\Delta$ that counts
the inverse powers of the large mass scale $M_{QCD}$,
\begin{equation}
\Delta = n + m + q+ 2p +\frac{f}{2} -2 \equiv d+\frac{f}{2} -2.
\label{index}
\end{equation}
Because of chiral symmetry, there are no interactions with
negative chiral index.
We can then split the chiral Lagrangian in pieces
${\cal L}^{(\Delta)}$:
\begin{equation}
\sum_{\Delta = 0}^\infty {\cal L}^{(\Delta)} = {\cal L}^{(0)} + {\cal L}^{(1)}
+{\cal L}^{(2)} + \ldots
\end{equation}
The form of the Lagrangian depends on the parametrization
of the pion fields.
For example, using stereographic coordinates, 
the lower-order Lagrangians are 
\cite{ciOvK,ciOLvKb,civK1}
\begin{eqnarray}
{\cal L}^{(0)} & = & -2f_\pi^2 \boldD_\mu^2
         - \frac{m_{\pi}^{2}}{2} D^{-1}\boldpi^{2} 
          +N^\dagger i{\cal D}_{0} N 
         -2g_{A}\vec{\boldD} \cdot N^\dagger\boldt\vec{\sigma} N  
\nonumber  \\
&& + C_0^{(S)} N^\dagger N \; N^\dagger N 
   +C_0^{(T)}  N^\dagger \vec{\sigma} N \cdot N^\dagger \vec{\sigma} N
\nonumber  \\
   &   & +\Delta^\dagger (i{\cal D}_{0}- \delta)\Delta 
         -2h_{A} \vec{\boldD}\cdot
          \left[N^\dagger \boldT\vec{S}\Delta+h.c.\right]
+\ldots, \label{L0}
\end{eqnarray}
\begin{eqnarray}
 {\cal L}^{(1)} & = & \frac{1}{2m_N} N^\dagger \vec{\cal D}^{\, 2} N
                     -\frac{g_A}{2m_N f_\pi} \boldD_0 \cdot 
  \left[iN^\dagger \boldt \vec{\sigma}\cdot \vec{\cal D}N
        + h.c.\right] \nonumber  \\
  & &   -B_{1}\boldD_\mu^2  N^\dagger N 
      -B_{2} (\vec{\boldD}\times\vec{\boldD})
            \cdot N^\dagger \boldt\vec{\sigma}N 
   -\frac{B_{3}
        m_{\pi}^{2}}{4f_{\pi}^{2}}D^{-1}\boldpi^{2}N^\dagger N
     -B_4 \boldD_0^2  N^\dagger N  \nonumber  \\
 & &  -D_1 \vec{\boldD} \cdot N^\dagger\boldt\vec{\sigma} N \;
           N^\dagger N 
       -D_2 \vec{\boldD}\cdot
           \left(N^\dagger\boldt\vec{\sigma}N 
                 \times N^\dagger\boldt\vec{\sigma}N \right)
       \nonumber  \\
 &&   -\frac{E_1}{2} N^\dagger N \; 
               N^\dagger\boldt N \cdot N^\dagger\boldt N 
      -\frac{E_2}{2} N^\dagger N \; 
               N^\dagger\boldt\vec{\sigma}N \cdot N^\dagger\boldt\vec{\sigma}N 
       \nonumber  \\
 &&         -\frac{E_3}{2} N^\dagger\boldt\vec{\sigma}N \cdot
                 \left(N^\dagger\boldt\vec{\sigma}N 
                 \times N^\dagger\boldt\vec{\sigma}N \right)
 \nonumber  \\
  & & + \frac{1}{2m_N} \Delta^\dagger \vec{\cal D}^{\, 2} \Delta
  -\frac{h_A}{2m_N f_\pi}\boldD_0 \cdot 
 \left[iN^\dagger \boldT \vec{S}\cdot \vec{\cal D}\Delta
         + h.c.\right]       +\ldots 
\label{L1}
\end{eqnarray}
\noindent
Here $g_{A}$, $h_A ={\cal O}(1)$ and $B_{i}={\cal O}(1/M_{QCD})$
are undetermined
constants, to be obtained either by solving QCD or by fitting data;
``\ldots'' stand for other terms involving the delta isobar.
Higher-index interactions can be constructed similarly.

\subsection{Why is nuclear physics so interesting?}

Processes that involve at most one nucleon ($A\le 1$)
can be easily described in the EFT.
If all external momenta are of the same order $Q\sim M_{nuc}$,
there are only two scales $Q$ and $M_{QCD}$.
A generic contribution to an amplitude can be written as
\begin{equation}
T \sim {\cal N} \left(\frac{Q}{M_{QCD}}\right)^\nu 
        {\cal F}\left(\frac{Q}{m_\pi}\right),
\label{form}
\end{equation}
where ${\cal F}$ is a dimensionless non-analytic function and
${\cal N}$ is a normalization factor.
Counting powers of $Q$ in a particular diagram
can be done as for the superficial degree
of divergence.
For a diagram with 
$L$ loops and
$V_\Delta$ vertices of index $\Delta$ we find \cite{weinberg79}
\begin{equation}
\nu=2- A + 2L+{\sum _\Delta}{V_\Delta}{\Delta}.
\label{pc}
\end{equation}
This formula is important because chiral symmetry places a lower
bound on the interaction index ${\Delta}\geq 0$.
Since $L$ is bounded from below ($L\ge 0$),
$\nu\ge \nu_{min}=2-A$ for strong interactions.
An expansion in  $Q/M_{QCD}$ results.
It starts at $\nu= \nu_{min}$ with tree ($L=0$) diagrams
built out of vertices of index 0 ($\sum V_\Delta \Delta =0$),
then proceeds at $\nu= \nu_{min}+1$ with further tree diagrams,
now with one vertex of index 1, the remaining
having index 0 ($\sum V_\Delta \Delta =1$). 
These first two orders are equivalent
to the current algebra of the 1960's, but now 
unitarity corrections can be accounted for systematically.
At $\nu= \nu_{min}+2$, for example, besides tree
diagrams with one index-2 interaction or two 
index-1 interactions ($\sum V_\Delta \Delta =2$),
there are also one-loop ($L=1$) diagrams built out of
index-0 vertices ($\sum V_\Delta \Delta =0$).
This is generalized to higher orders in obvious fashion.
An estimate of the expansion parameter is 
$\sim m_\pi/m_\rho \approx 0.2$.
In this context the EFT is called Chiral Perturbation Theory (ChPT).
See Ref. \cite{bkm} for a review.

\begin{figure}[t]
\includegraphics[totalheight=1.25in]{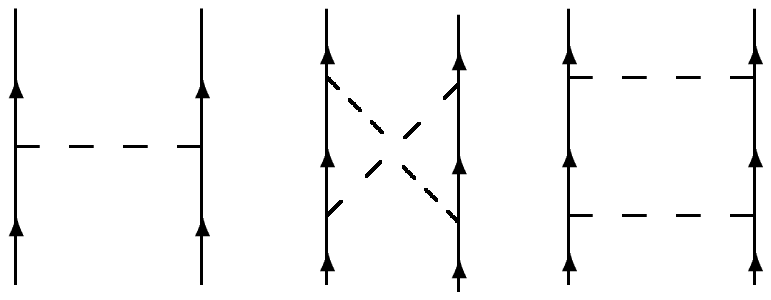}
\caption{Simplest pion-exchange diagrams in the $NN$ amplitude.
\label{pionex}}
\end{figure}

Processes with two or more nucleons ($A\ge 2$) are much more interesting.
One can see this by studying the simplest Feynman diagrams 
contributing to the $NN$ amplitude $T_{NN}$, shown in Fig. \ref{pionex}.
The first diagram results in the term:
\begin{equation}\label{ope}
T_{NN}^{(1)}=\frac{g_A^2}{f_\pi^2} \frac{\vec{\sigma}_1\cdot \vec{q}
\vec{\sigma}_2\cdot \vec{q}}{q^2 + m_\pi^2} \sim \frac{g_A^2}{f_\pi^2}.
\end{equation}
This is analogous to Coulomb-photon exchange in QED.
By following the same steps as in that example,
we see that the crossed-box diagram is
\begin{equation}
T_{NN}^{(2)(\times)} \sim \frac{g_A^4}{f_\pi^4}
\Big(\frac{Q^3}{4\pi}\Big)\Big(\frac{1}{Q}\Big)\Big(\frac{1}{Q}\Big)
\Big(\frac{Q^2}{Q^2}\Big)\Big(\frac{Q^2}{Q}\Big)
\sim \frac{g_A^2 Q^2}{4\pi f_\pi^2} \frac{g_A^2}{f_\pi^2},
\end{equation}
which is small, given that $4\pi f_\pi \sim M_{QCD}$.

However, the box diagram is
\begin{equation}
T_{NN}^{(2)(x)}\sim \frac{g_A^4}{f_\pi^4}
\Big(\frac{Q^3}{4\pi}\Big)\Big(\frac{m_N}{Q^2}\Big)
\Big(\frac{Q^2}{Q^2}\Big)\Big(\frac{Q^2}{Q^2}\Big)
+\ldots
\sim \frac{g_A^2 m_N Q}{4\pi f_\pi^2} \frac{g_A^2}{f_\pi^2}
+\ldots
\end{equation}
The scale that sets the relative size of this diagram
is not $M_{QCD}$ but 
\begin{equation}
M_{NN}\equiv \frac{4\pi f_\pi^2}{g_A^2 m_N},
\end{equation}
which numerically is $\sim f_\pi \sim M_{nuc} \ll M_{QCD}$.

As in the QED case, the infrared enhancement 
of $m_N/Q$ over irreducible states appears in
all reducible intermediate states.
Contrary to ordinary ChPT, we need to resum
the leading terms; in a schematic way,
\begin{eqnarray} \label{TNN}
T_{NN}
&\sim&\frac{g_A^2}{f_\pi^2}\left[1+{\cal O}\left(\frac{Q}{M_{NN}}\right)
     +\cdots+ {\cal O}\left(\frac{Q}{M_{QCD}}\right) \right] \nonumber \\
&\sim&\frac{g_A^2}{f_\pi^2}\left[\frac{1}
       {1-{\cal O}\left(\frac{Q}{M_{NN}}\right)}
     +\cdots+ {\cal O}\left(\frac{Q}{M_{QCD}}\right)\right].
\end{eqnarray}
Nuclear bound states thus appear
at 
\begin{equation}
Q\sim M_{NN},
\end{equation}
with binding energy
\begin{equation}
B\sim\frac{Q^2}{m_N}\sim \frac{M_{NN}^2}{m_N}.
\end{equation}
Numerically this is $\sim M_{QCD}/(4\pi)^2 \approx 10$ MeV, which explains why
nuclei are so shallow when compared to the characteristic QCD
scale.

\subsection{Deriving a potential for nuclear physics}

We can now carry out systematic calculations
by finding a power counting for the potential,
defined as the sum of all irreducible diagrams,
and then solving the Lippman-Schwinger equation
(Fig. \ref{potential}) order by order.

Weinberg \cite{inwei6} was the first to suggest a simple power counting. 
He reasoned that, because
the enhanced states are by construction removed from the potential,
powers of $Q$ could be counted in the potential as in ChPT.
This results in an expansion of the form:
\begin{equation}
 V = \sum_{\nu = 0}^\infty c_\nu Q^\nu,
\end{equation}
where
\begin{equation}\label{nu}
\nu = 4-A + 2 (L -C) + \sum_\Delta V_\Delta \Delta. 
\end{equation}
The only difference compared to Eq. (\ref{pc}) is the number of separately
connected pieces $C\ge 1$. They arise because the potential is just
part of the full amplitude and does not need to be fully connected.

\begin{figure}[t]
\includegraphics[totalheight=5.0in]{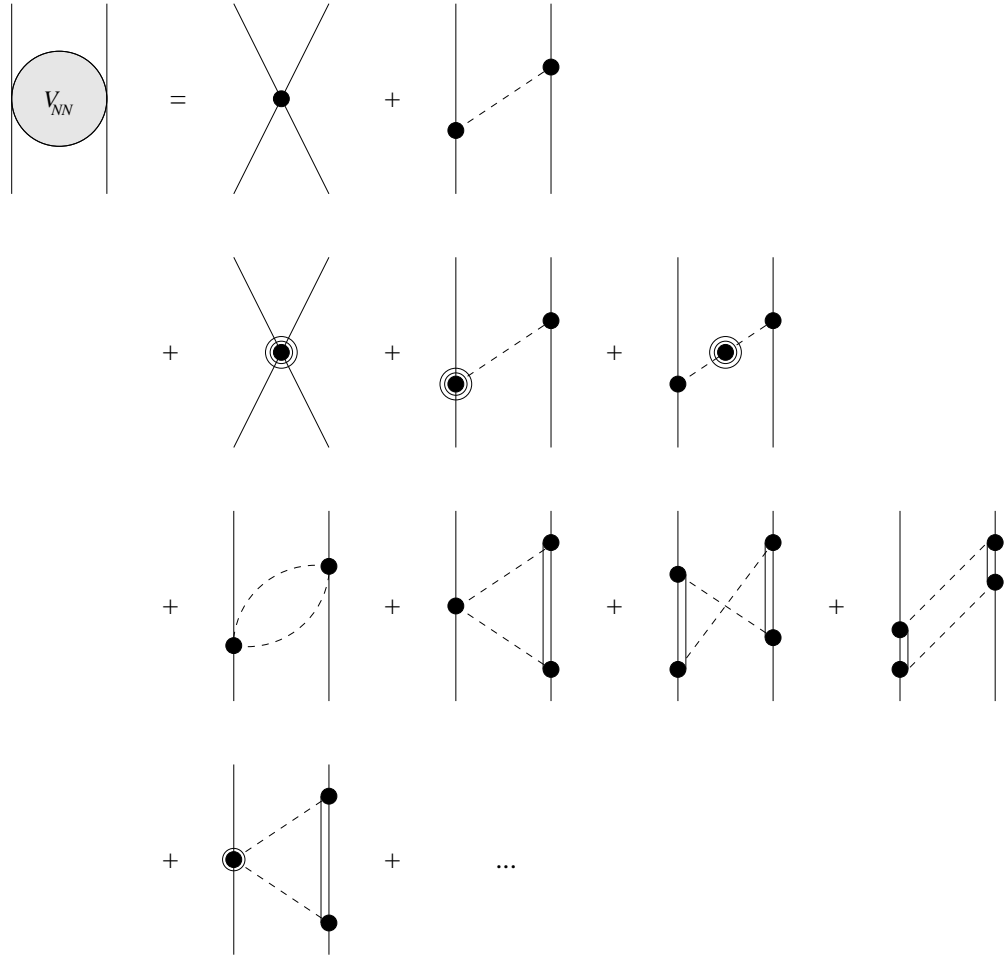}
\caption{Some time-ordered diagrams contributing to the two-nucleon potential
$V_{NN}$ in the EFT.
(Double) solid lines represent nucleons (and/or deltas), 
dashed lines pions, a heavy dot an interaction in ${\cal L}^{(0)}$,
a dot within a circle an interaction in ${\cal L}^{(1)}$,
and 
a dot within two circles an interaction in ${\cal L}^{(2)}$.
First line corresponds to $\nu=\nu_{min}$,
second and third lines to $\nu=\nu_{min}+2$, 
fourth line to $\nu=\nu_{min}+3$, and 
``$\ldots$'' denote $\nu\geq \nu_{min}+4$. 
All orderings with at least one pion
or delta in intermediate states are included.
Not shown are 
diagrams contributing only to renormalization of parameters.
\label{F:vkolck:V2N}}
\end{figure}

According to this power counting,
in leading order we have, besides one-pion exchange (OPE)
between two nucleons, also
non-derivative two-nucleon contact interactions 
(the $C_0^{(S)}$ and $C_0^{(T)}$ terms in Eq. (\ref{L0}).)
A calculation of all (isospin-conserving) contributions to the
two-nucleon potential up to $\nu=\nu_{min}+3$ was carried out
in Refs. \cite{ciOvK,ciOLvKb}
using time-ordered perturbation theory. 
Some diagrams are shown in Fig. \ref{F:vkolck:V2N}.
To this order,
the potential has all the spin-isospin structure of the phenomenological
models, but its profile is determined by explicit degrees of freedom,
symmetries, and power counting. 
The power counting suggests a hierarchy of short-range effects:
$S$ waves should depend strongly on the short-range parameters $C_0^{(S,T)}$;
contact interactions affect $P$-wave phase shifts 
only in subleading order,
so their effect should be smaller and approximately linear;
$D$ waves are affected by contact interactions only via mixing,
while higher waves should be essentially determined by pion exchange.
Chiral symmetry is particularly influential
in the two-pion exchange (TPE) piece. 
The latter includes a particular form of terms previously
considered \cite{ciBw}, plus a few new terms.
Those terms involving the $B_i$'s and the deltas provide the only form of
correlated TPE to this order, as graphs where pions interact in flight appear
only in next order and should thus be relatively small. 
The sum of the $B_1$ term and the corresponding
delta term (related to the nucleon axial polarizability),
is particularly important in providing an isoscalar central force.
Not surprisingly, in the chiral limit these potentials 
behave at large separations as van der Waals forces.
Isospin-violating pieces of the potential can be calculated as well
\cite{iv}.

\begin{figure}[t]
\includegraphics[totalheight=5.0in]{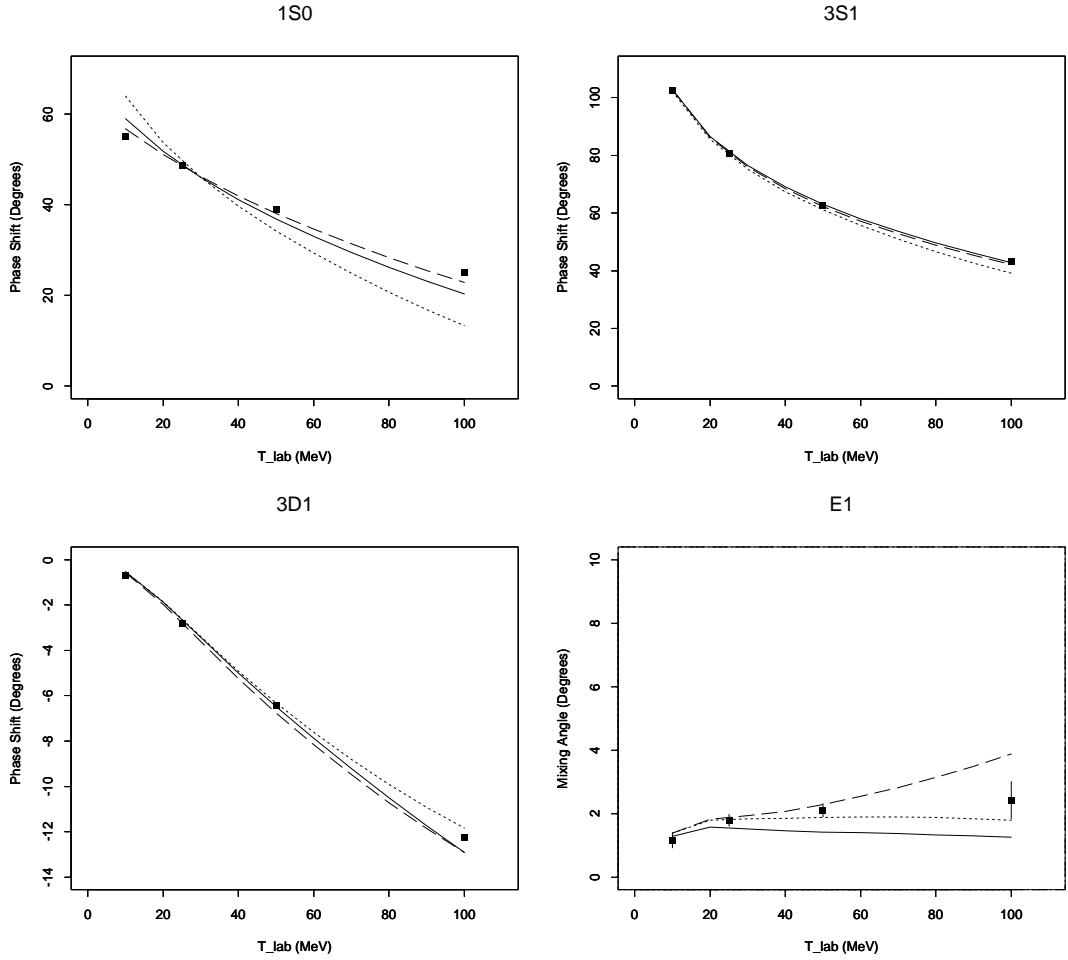}
\caption{ $^1S_0$, $^3S_1$, and $^3D_1$ $NN$ phase shifts and 
$\varepsilon_1$ mixing angle in degrees as functions
of the laboratory energy in MeV: 
EFT up to $\nu=3$ for cutoffs of 
500 (dotted), 780 (dashed), and 1000 MeV (solid line); 
and Nijmegen PSA (squares).}
\label{F:vkolck:NN}
\end{figure}

\begin{figure}
\includegraphics[totalheight=3.0in]{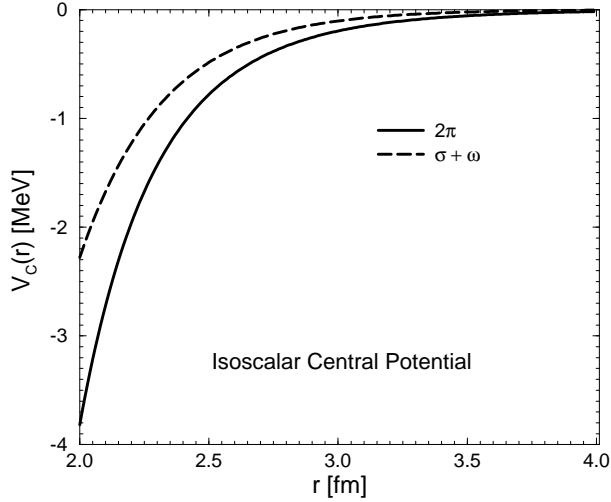}
\caption{\label{sigma} 
The isoscalar central potential generated by two-pion exchange
compared with phenomenological $\sigma+\omega$ contributions~\cite{sigma}.}
\end{figure}

Using this
potential, one can then solve the Schr\"odinger equation.
For each cutoff the parameters are fitted
to data at selected energies.
A sample of the results \cite{ciOLvKb}
for the lowest, most important partial waves
is presented in Fig. \ref{F:vkolck:NN},
together with phases from the Nijmegen phase-shift analysis \cite{nijmanal}.
Quality of the fits is typical of other waves.
Waves higher than $F$-waves were found to be mostly described
well by pion exchange alone, as expected. 
Deuteron quantities 
are shown in Table \ref{tab:dparam}.
Electromagnetic quantities refer to the contributions
from lowest-order $\gamma NN$ couplings only,
not to a consistent calculation which would include
sub-leading one- and two-nucleon effects.
The predicted $S$-wave scattering lengths 
(not used to constrain the fit) were found to be
$a_2^{(^1S_0)}\simeq -15.0$ fm and 
$a_2^{(^3S_1)}\simeq 5.46$ fm.
Important central attraction comes from the $B_i$'s and deltas,
and indeed the central potential does resemble those from 
models that include $\sigma$ and $\omega$ meson 
exchange explicitly \cite{sigma}
(see Fig. \ref{sigma}). Thus the properties of the mesons
extracted from phenomenological models have limited meaning.
Values for the parameters are listed in Ref. \cite{ciOLvKb}.
Reasonable values were found for quantities known at the time,
for example 
$g_A=1.33$ (in agreement with the Goldberger-Treiman relation)
and $h_A=2.03$ (smaller but not too far off the large-${\cal N}_c$ value).
However, the values for the $B_i$'s came out somewhat different than
those found later in $\pi N$ scattering. 
For a cutoff of $\Lambda=780$ MeV, 
the coefficients $C_{2n}$ of the contact interactions
were found to scale approximately 
with $M_{QCD} \sim 500$ MeV.

\begin{table}[b]
\caption{Results from EFT fits at $\nu=3$ for various cut-offs
$\Lambda$ and experimental values for the deuteron binding energy
($B$), magnetic dipole moment ($\mu_d$), electric quadrupole moment
($Q_E$), asymptotic $D/S$ ratio ($\eta$), and $D$-state probability
($P_D$).\label{tab:dparam}}
\vspace{0cm}
\footnotesize
\begin{tabular}{|ccccc|}
\hline
Deuteron   & \multicolumn{3}{c}{$\Lambda$ (MeV)} &            \\
\cline{2-4}
quantities & 500 & 780 & 1000                   & Experiment \\
\hline
$B$ (MeV)  & 2.15 & 2.24 & 2.18                 & 2.224579(9) \\
$\mu_d$ ($\mu_N$) & 0.863 & 0.863 & 0.866       & 0.857406(1) \\
$Q_E$ (fm$^2$) & 0.246 & 0.249 & 0.237          & 0.2859(3)   \\
$\eta$     & 0.0229 & 0.0244 & 0.0230           & 0.0271(4)  \\
$P_D$ (\%) & 2.98 & 2.86 & 2.40                 &            \\
\hline
\end{tabular}
\end{table}

The EFT also offers some insight into few-nucleon forces.
Weinberg's power counting embodied
in Eq. (\ref{nu}) suggests a hierarchy of few-nucleon forces.
In leading order ($\nu=\nu_{min}=6-3A$), $C$ is maximum, so we
have only pairwise interactions via 
the leading two-nucleon potential.
We can easily verify that, if the delta is kept
as an explicit degree of freedom, a $3N$ potential will arise at 
$\nu=\nu_{min}+2$,
a $4N$ potential at $\nu=\nu_{min}+4$, and so on. 
It is (approximate) chiral symmetry therefore that
implies that $n$-nucleon forces $V_{nN}$ obey a hierarchy
of the type
\begin{equation}
\frac{\langle V_{(n+1)N}\rangle}{\langle V_{nN}\rangle}
\sim {\cal O}\left( \frac{Q}{M_{QCD}} \right)^2,
\end{equation}
with $\langle V_{nN}\rangle$ denoting the contribution 
per $n$-plet.
If 
$|\langle V_{2N}\rangle | \sim M_{QCD}/(4\pi)^2 \approx 10$ MeV,
we can expect
$|\langle V_{3N}\rangle | \sim 0.5$ MeV,
$|\langle V_{4N}\rangle | \sim 0.02$ MeV, and so on. 
This is in accord with detailed few-nucleon phenomenology
based on potentials that include small $3N$ and no $4N$ forces.
This is shown in Table \ref{table:sizes}
in the case of the AV18/IL2 potential \cite{Pieper:2001mp}.

\begin{table}[!b]
\caption{Contributions of the two-, three- and four-nucleon potentials
(per doublet, per triplet, and per quadruplet, respectively):
Weinberg power counting (W pc) and calculations 
with the AV18/IL2 potential
for the ground states of various light nuclei ($^2$H, $^3$H, {\it etc.}).
\label{table:sizes}}
\vspace{0.25cm}
\begin{tabular}{||c||c||c|c|c|c|c|c|c|c||} 
\hline
(MeV)  & W pc & $^2$H &  $^3$H & $^4$He & $^6$He & $^7$Li & $^8$Be & $^9$Be 
       & $^{10}$B \\
\hline
$|\langle V_{2N}\rangle |$& $\sim 10$ & 22 & 20 & 23 & 13 & 11 & 11 &  9.4 
                          & 8.9 \\
$|\langle V_{3N}\rangle |$& $\sim 0.5$ & -- & 1.5 & 2.1 & 0.55 & 0.43 & 0.38 & 0.29 
                          & 0.30 \\
$|\langle V_{4N}\rangle |$& $\sim 0.02$& -- & -- & ? & ? & ? & ? & ? 
                          & ?\\
\hline
\hline
{ $|\langle V_{3N}\rangle |/|\langle V_{2N}\rangle |$}
& $\sim 0.05$ & -- &0.075 & 0.091 & 0.042 & 0.039 & 0.035 &  0.031 
& 0.034\\

{ $|\langle V_{4N}\rangle |/|\langle V_{3N}\rangle |$}
& $\sim 0.05$ & -- & -- & ? & ? & ? & ? & ? 
& ? \\
\hline
\end{tabular}
\end{table}

\begin{figure}[!t]
\includegraphics[totalheight=2.25in]{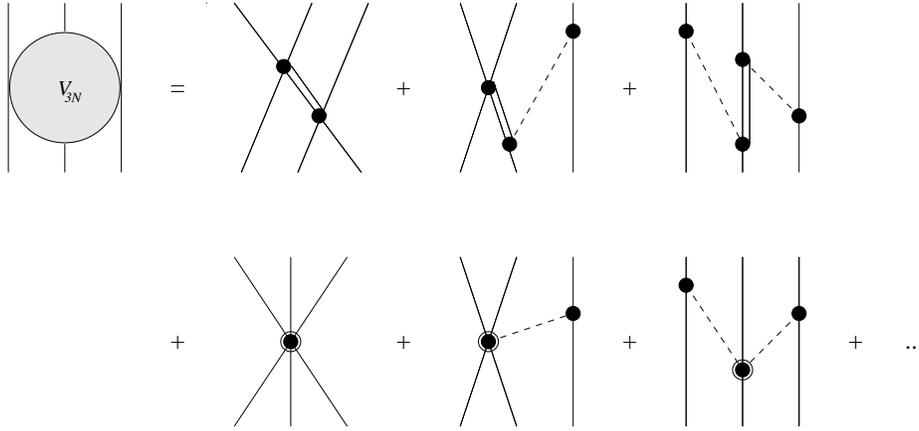}
\caption{Some time-ordered diagrams contributing to the $3N$ potential
in the pionful EFT.
(Double) solid lines represent nucleons (and/or deltas), 
dashed lines pions, a heavy dot an interaction in ${\cal L}^{(0)}$,
and a dot within a circle an interaction in ${\cal L}^{(1)}$.
First line corresponds to $\nu=\nu_{min}+2$,
second line to $\nu=\nu_{min}+3$, 
and ``$\ldots$'' denote $\nu\geq \nu_{min}+4$. 
All nucleon permutations and orderings with at least one pion
or delta in intermediate states are included.}
\label{F:vkolck:V3N}
\end{figure}

The new forces that appear in systems with more than two 
nucleons have been derived in 
Refs. \cite{ciOvK,civK1}.
The relevant terms up to $\nu_{min}+3$ are shown
in Fig. \ref{F:vkolck:V3N}.
The leading $3N$ potential 
has components with three different ranges:
TPE; OPE/short; and purely
short.
The TPE part of the potential is determined
in terms of $\pi N$ scattering observables,
and is similar to existing two-pion-exchange $3N$ potentials
\cite{vk:huber}.
The novel OPE/short-range
components 
involve $\pi (N^\dagger N)^2$ interactions of strengths
that are not fixed by chiral symmetry alone but
can be determined 
from reactions involving only two nucleons,
such as $NN\rightarrow NN\pi$ \cite{huber2}.
The purely short-range components of the potential can only be determined from
few-nucleon systems.

The nuclear potential from EFT has been further elaborated in the 
last couple of years. 
Calculations are being pushed to next order \cite{kaiserloops},
better fits (with more extensive input from
$\pi N$ scattering) to $NN$ data have been achieved \cite{epelfit},
and the first results for $3N$ and $4N$ systems have appeared \cite{epel34NLO}.
These developments are all reviewed in Ref. \cite{vankolck99}.

\begin{figure}[t]
\includegraphics[totalheight=3.0in]{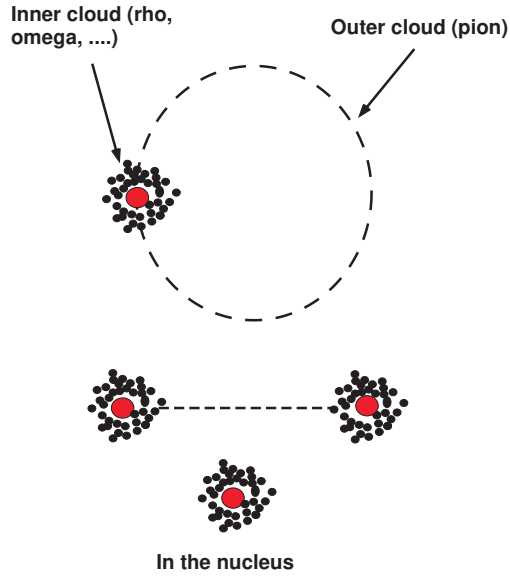}
\caption{The short-ranged ($\rho$, etc.) cloud around the nucleon and the
long-ranged ($\pi$) cloud surrounding it.
\label{nucleon}}
\end{figure}

Weinberg's power counting thus provides a rationale to
understand much of the phenomenology of light nuclei.
At low energies, the nucleon can be visualized as in Fig. \ref{nucleon}:
the light pion states form 
a sparse outer cloud, leading to a loop expansion in
$Q/4\pi F_\pi$, while the 
high-energy states form a
dense inner cloud, amenable to
a multipole expansion in terms
of $Q/m_\rho$. 
Since most of the time there is only a single pion ``in the air'',
in the nucleus the interaction among nucleons is mostly pairwise,
resulting also in a cluster expansion for the potential.

\subsection{Renormalization}

Despite all similarities, there is an important difference between
the nuclear and atomic potentials.
Because of chiral symmetry, the potential from one-pion
exchange goes as $Q^0$, from two-pion exchange as $Q^2$, and so on.
The loops generated by the iteration of these potentials
are then sensitive to the cutoff.
In coordinate-space parlance, these potentials
are singular, that is, they behave for small radial distances $r$
as $1/r^3$, $1/r^5$, and so on.
The Schr\"odinger equation for such potentials
is not well defined.
Clearly, one cannot consider pion exchange without 
including enough short-distance
operators to erase the cutoff dependence from observables.

Now, Weinberg's power counting automatically predicts
which contact operators have to be taken at each order.
Following this prescription, it does seem that 
the fits are of the same quality for different cutoffs.
However, the numerical character of the calculations
makes it hard to offer a proof of consistency.
It is not apparent, for example, how a momentum-independent 
short-range interaction can renormalize 
the OPE potential (\ref{ope}).

Much effort has been devoted to this issue in the last five years
\cite{vankolck99}.
A search for more analytical approaches led to extensive
studies of the EFT at momenta $Q \ll m_\pi$.
In this regime, pions can be treated as heavy particles.
All interactions are of contact type,
and form an expansion in $Q/M_{nuc}$.
This ``pionless'' EFT can be solved by hand in the $NN$ sector,
and by very simple numerical means in the $3N$ and $4N$ systems.
Power counting and renormalization are not without surprises,
but they can be done consistently.
Of course, the pionless EFT 
does not address the problem of pion exchange directly,
but it goes a long way in elucidating how
EFTs work in a non-perturbative context.
Moreover, it is useful for many low-energy reactions, to which it
has by now been applied.

As $Q$ is increased, the approximation of zero-range propagation
for pions becomes less and less reliable. 
At momenta $Q\sim m_\pi$, pions have to be included explicitly in the
theory. 
Once pions have been reinstated,
we can still consider the low-$Q$ region.
Because of chiral symmetry, it is reasonable 
to suppose that pion interactions are perturbative
at sufficiently low $Q$.
Based on the power counting of the pionless EFT, a
new power counting was formulated that led to
a manifestly consistent EFT where pion exchange was treated in
perturbation theory \cite{ksw_1}.
Unfortunately, it has been shown that the range
of validity of this power counting is $Q\saprox 100$ MeV \cite{fms}.
In this region the simplest pionless EFT is more useful.

Recently, the non-perturbative renormalization
of pion exchange was analyzed \cite{towards}.
It was found that OPE can be renormalized
\`a la Weinberg, provided one expands in the pion mass
even the pion propagators.
If one does not do expand, one makes numerically small
but conceptually large errors.
The full implication of this conclusion to previous results
remains to be explored.

\subsection{Nuclear matter on a lattice} 

The nuclear EFT that has been sketched above can be used to
study nuclear matter at finite temperature. This provides a link
with RHIC physics that is the subject of this meeting. 

The general concept of a nuclear matter calculation consists of
nucleons interacting via a variety of components of the nuclear
potential. While the ultimate goal is to use EFT interactions, 
let us 
concentrate for simplicity on few parts of the $NN$ potential, namely 
central, spin- and isospin-exchange. 
The Hamiltonian,
\begin{equation}
\hat{\cal H} = \hat{\cal K} + \hat{\cal V},
\end{equation}
can be expressed in second quantization and contains kinetic and
potential operators. The kinetic term is written as
\begin{equation}
\hat{\cal K} = - \psi^\dag \frac{\nabla^2}{2 m_N} \psi,
\end{equation}
while the potential is taken as
\begin{equation}
\hat{\cal V} =  C_0 \L(\psi^\dag  \psi\R)^2 
             + C_2 \L(\psi^\dag \nabla  \psi\R)^2.
\end{equation}
The fermion operator $\psi^\dag$ creates
a nucleon of spin and isospin $(\sigma, \tau)$ at location
$\vec{x}$, while its adjoint $\psi$
destroys it. This Hamiltonian describes merely a toy model
for developing the formalism. More complicated Hamiltonians arising
from full-fledged nuclear EFT can be later used for more
realistic calculations.

How can we deal with the complication of many nucleons?
A natural approach, patterned after similar
attempts in QCD, is to investigate nuclear 
matter on a
three-dimensional cubic lattice of
spacing $a$ and periodic boundary conditions. 
We 
describe the nuclear-matter Monte Carlo method \cite{Muller:1999cp}, 
which consists of the thermal formalism to express the
grand-canonical partition function as an integral over single-body
evolution operators. At its center stands the
Hubbard-Stratonovitch transformation, which is used to reduce the
many-body problem to an effective one-body problem. 

In order to study thermal properties of nuclear matter, the 
grand-canonical partition function at a given temperature $T=\beta^{-1}$
needs to be determined:
\begin{equation}
Z = {\rm \hat{Tr}} \left[ \exp \left( -\beta \left( \hat{\cal H} -
\psi^\dag \mu \psi \right)
\right) \right] \equiv {\rm \hat{Tr}} \left[ \hat{U} \right],
\end{equation}
with $\mu$ as the chemical
potential. $\hat{U}$ is called the imaginary-time evolution
operator of the system and is a many-body operator; the trace is
taken over all many-body states. 
The
partition function $Z$ is an exponential over all one- and
two-body operators (and therefore interactions) present in the
system. It is impossible to deal with $Z$
in this form, because the number of many-body correlations that
have to be kept track of grows rapidly with system size. We
therefore find an expression for $Z$ that is based on a
single-particle representation, replacing the many-body
problem with that of non-interacting nucleons that are coupled to
a heat bath of auxiliary fields. 
A thermal observable $\langle \hat{O} \rangle$ is then expressed as
\cite{gubernatis}
\begin{equation}
\label{therobs} \langle \hat{O}\rangle = \frac{1}{Z} {\rm
\hat{Tr}} \left[\hat{O} \exp \left( -\beta \left(\hat{{\cal H}}-
\psi^\dag \mu \psi \right)
\right)\right] = \frac {\int {\cal D} [ \chi ] G(\chi) \langle
\hat{O}(\chi) \rangle \xi(\chi)}{\int {\cal D} [ \chi ]
G(\chi)\xi(\chi)},
\end{equation}
where the Gaussian factor $G$ is given by
\begin{equation}
\label{gaussfac} G\left(\chi\right) = \prod_{m=1}^{n_t}
\prod_{\vec{x}_n} \prod_{i} \exp \left( - |\alpha_i| \chi^2_{m,
\vec{x}_n, i} \right).
\end{equation}
The integrals in this equation can then be evaluated using the
Metropolis algorithm \cite{metropolis} (Monte Carlo simulations).

\begin{figure}
\includegraphics[totalheight=2.8in,angle=-270]{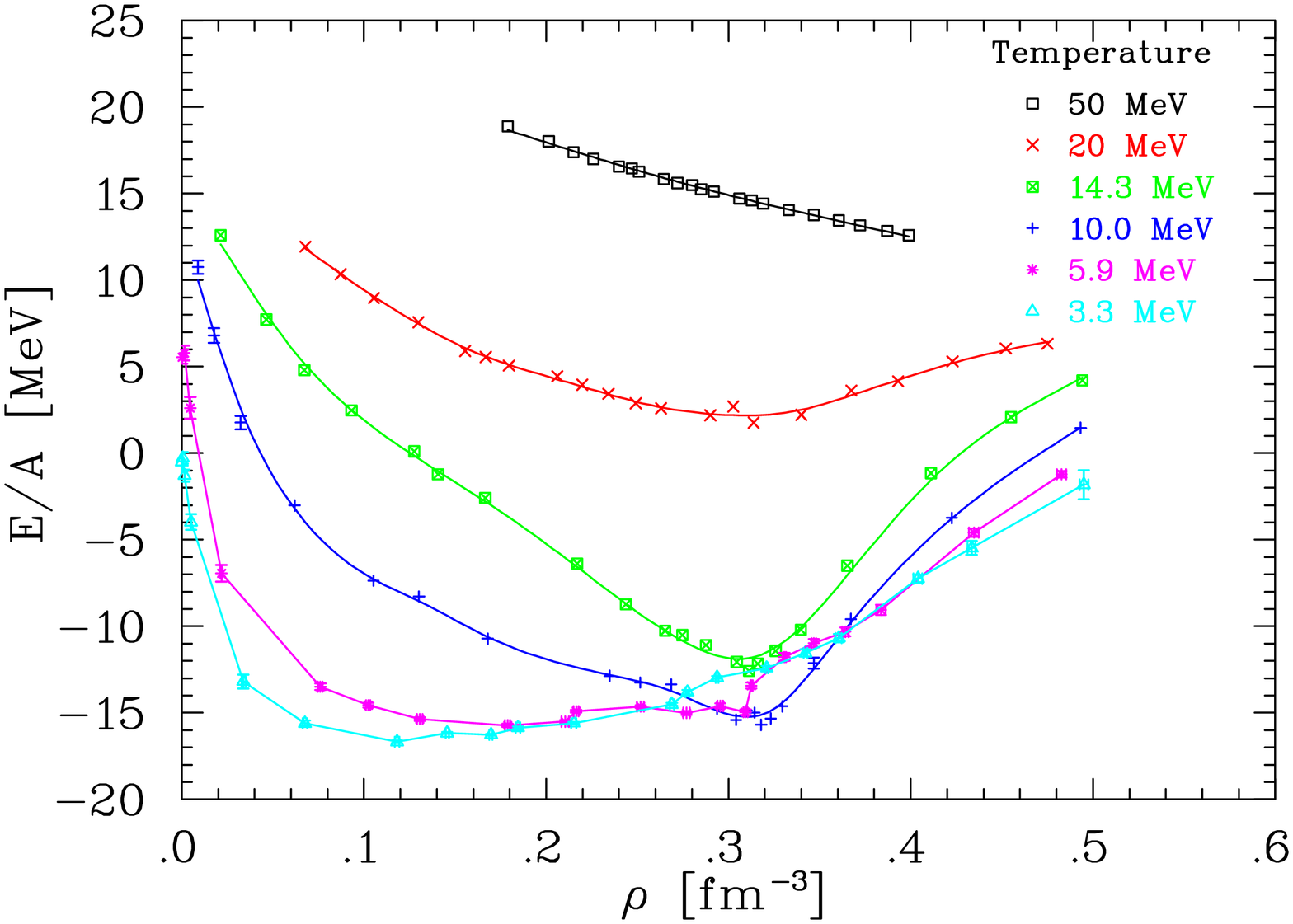}
\includegraphics[totalheight=2.8in,angle=-270]{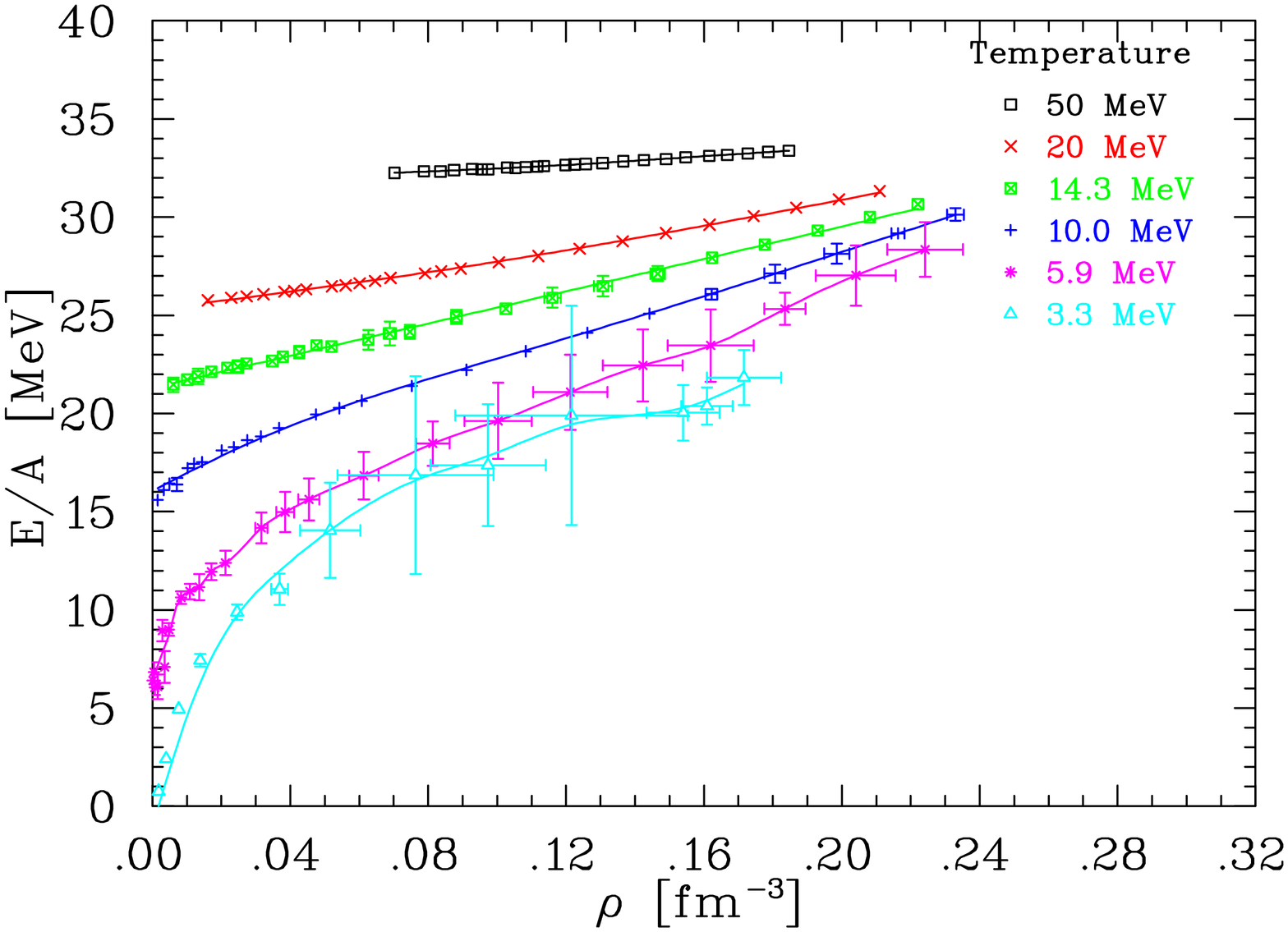}
\caption{$E/A$ for symmetric nuclear (left)
and pure neutron (right) matter as a function of
density $\rho$ and for different temperatures. The purpose of the
lines is to guide the eye.} \label{snmenergy}
\end{figure}

Fig. \ref{snmenergy} \cite{Muller:1999cp} shows that such calculations 
are feasible.
It displays the energy per particle $E/A$ for symmetric nuclear matter 
and for pure
neutron matter as a function of
density $\rho$ and for different temperatures. 
The two interaction parameters were adjusted so that
the known {\it qualitative} features of the symmetric nuclear matter
are reproduced.
With
decreasing temperature, symmetric nuclear matter 
develops a minimum at $\rho =
0.32\; {\rm fm^{-3}}$ first, which is most pronounced between
$10-14 \; {\rm MeV}$, before it shifts to lower densities. At
$T=3.3 \; {\rm MeV}$ and $T=5.9 \; {\rm MeV}$ the minimum is very
broad, making matter softer. For high temperatures and/or high
density, the simulation suffers from the fact that it runs out of
model space. At $T=50\; {\rm MeV}$ the system behaves almost like
a Fermi gas and the energy per particle should behave like $\sim
\rho^{2/3}$. Yet, the curve bends down. Also, for all other
temperatures, the curves converge to the energy of the full
lattice state, $E/A = 5.96 \; {\rm MeV}$, as density increases.
For sub-saturation densities the model gives more binding if
compared to other calculations (see, for example, Ref. \cite{wff}), 
and the energy is not as high for densities
beyond saturation. At $\rho = 0.32 \; {\rm fm^{-3}}$, $E/A$ as a
function of temperature has a minimum at $T \approx 10 \; {\rm
MeV}$ which means that at even lower temperatures $E/A$ increases
again. We see some evidence for a liquid-gas phase transition. 
The uncertainties for pure neutron matter are much larger
than for symmetrical nuclear matter. As a potential, we used the
parameters obtained from the fit to symmetric nuclear matter, even
though we could have fitted the potential parameters for this case
anew. Therefore we view
the results for pure neutron matter more as a test to see how well
the given potential already reproduces the energy. Note that the
slopes of the curves at high temperatures are not negative as they are
for symmetric nuclear matter. But clearly, we cannot conclude
that the energies at $T= 3.3 \; {\rm MeV}$ have converged to that
of the ground state because the curve differs quite a bit from
that of $T = 5.9 \; {\rm MeV}$. At the lowest temperature they are
$4-5 \; {\rm MeV}$ higher than those of the ground state as
calculated in Ref. \cite{wff}, but the general shape of the curve
is very similar. This is no surprise, since pure neutron matter is
like a Fermi gas, with attractive forces between neutrons lowering
the energies with respect to the non-interacting system. The
search for any kind of phase transition in the range of $5-50 \;
{\rm MeV}$ was to no avail.

\section{Conclusion} \label{sec-out}

We have presented an introduction to EFT 
and some of the mileposts along the sinuous road
from QCD to nuclear physics.
We discussed some of the essential
features of the interactions among nucleons 
and some of their consequences
to few-nucleon systems.
We can already see how to approach the solution of
the important problem of finite-temperature nuclear matter. 
Yet, there is no denying that much is still to be done 
in straightening the path already traveled,
and in breaking new ground.
It is a good time to be on the road!

\begin{theacknowledgments}
The authors acknowledge the support and dedication of all PASI
sponsors and 
organizers,
in particular Jan Rafelski, Bob Thews, Thomas Elze, Erasmo Ferreira,
and Takeshi Kodama.
UvK thanks the Nuclear Theory Group at the University of Washington
for hospitality while part of this work was carried out,
and RIKEN, Brookhaven National Laboratory and to the U.S.
Department of Energy [DE-AC02-98CH10886] for providing the facilities
essential for the completion of this work.
This research was supported in part 
by a DOE Outstanding Junior Investigator Award (UvK).
\end{theacknowledgments}

\bibliographystyle{aipproc}   
\bibliography{eft}
\IfFileExists{\jobname.bbl}{}
 {\typeout{}
  \typeout{******************************************}
  \typeout{** Please run "bibtex \jobname" to obtain}
  \typeout{** the bibliography and then re-run LaTeX}
  \typeout{** twice to fix the references!}
  \typeout{******************************************}
  \typeout{}
 }
\end{document}